\documentclass{emulateapj}
\usepackage{graphicx}

\slugcomment{Accepted for publication in the ApJ}

\shorttitle{FG-Type Stars in SDSS DR3}
\shortauthors{Allende Prieto et al.}

\begin{document}

\title{A Spectroscopic Study of the Ancient Milky Way: F- and G-Type Stars in 
the Third Data Release of the Sloan Digital Sky Survey}

\author{Carlos Allende Prieto}
\affil{McDonald Observatory and Department of Astronomy, University of Texas,
    Austin, TX 78712}
\email{callende@astro.as.utexas.edu}    

\author{Timothy C. Beers}
\affil{Department of Physics \& Astronomy and JINA: Joint Institute for Nuclear
Astrophysics,\\ Michigan State University, E. Lansing, MI  48824}

\author{Ronald Wilhelm}
\affil{Department of Physics, Texas Tech University, Lubbock, TX 79409}

\author{Heidi Jo Newberg}
\affil{Rensselaer Polytechnical Institute, Troy, NY 12180}

\author{Constance M. Rockosi}
\affil{UCO/Lick Observatory, 1156 High St., Santa Cruz, CA 95064}

\author{Brian Yanny}
\affil{Fermi National Accelerator Laboratory, P.O. Box 500, Batavia, IL 60510}

\and

\author{Young Sun Lee}
\affil{Department of Physics \& Astronomy and JINA: Joint Institute for Nuclear
Astrophysics,\\ Michigan State University, E. Lansing, MI  48824}

~~

~~

~~

\begin{abstract}

We perform an analysis of spectra and photometry for 22,770 stars included in
the third data release (DR3) of the Sloan Digital Sky Survey (SDSS). We measure
radial velocities and, based on a model-atmosphere analysis, derive estimates of
the atmospheric parameters (effective temperature, surface gravity, and [Fe/H])
for each star. Stellar evolution models are then used to estimate distances. We
thoroughly check our analysis procedures using three recently published
spectroscopic libraries of nearby stars, and compare our results with those
obtained from alternative approaches. The SDSS sample covers a range in stellar
brightness of $14 < V < 22$, primarily at intermediate galactic latitudes, and
comprises large numbers of F- and G-type stars from the thick-disk and halo
populations (up to 100 kpc from the galactic plane),  therefore including
some of the oldest stars in the Milky Way. 

In agreement with previous results from the literature, we find that halo stars
exhibit a broad range of iron abundances, with a peak at [Fe/H] $\simeq -1.4$.
This population exhibits essentially no galactic rotation. Thick-disk G-dwarf 
stars at distances from the galactic plane in the range $1<|z|<3$ kpc 
show a much more compact metallicity distribution, with a maximum at [Fe/H] $\simeq -0.7$, and a
median galactic rotation velocity at that metallicity of $157 \pm 4$ km s$^{-1}$
(a lag relative to the thin disk of 63 km s$^{-1}$). 
SDSS DR3 includes spectra of many
F-type dwarfs and subgiants between 1 and 3 kpc from the plane with galactic
rotation velocities consistent with halo membership. A comparison of color
indices and metal abundances with isochrones indicates that no significant star
formation has taken place in the halo in the last $\sim 11$ Gyr, but there are
thick-disk stars which are at least 2 Gyr younger. We find the
metallicities of thick-disk stars to be nearly independent of galactocentric
distance between 5 and 14 kpc from the galactic center, in contrast with the
marked gradients found in the literature for the thin disk. No vertical
metallicity gradient is apparent for the thick disk, but we detect a gradient in
its rotational velocity of $-16 \pm 4$ km s$^{-1}$ kpc$^{-1}$ between 1 and 3
kpc from the plane. We estimate that among the stars in our sample there are
over 2000 with an iron abundance [Fe/H] $< -2$, and over 150 stars with an iron
abundance [Fe/H] $<-3$.

\end{abstract}


\section{Introduction}

Our present view of the structure of the Milky Way is mainly built upon star
counts. The most recent applications of this classical approach converge to
two-component disk models with scale heights of $Z^h_{\rm thin} \sim 0.2-0.3$
kpc and $Z^h_{\rm thick} \sim 0.6-1.0$ kpc, and a normalization factor
$\rho_{\rm thick}/\rho_{\rm thin} \sim 0.02-0.13$. Two more spheroidal
components are needed to account for the observed numbers of stars, a central
bulge and a halo. The density distribution of the stellar halo is still poorly
constrained, but it extends to distances in excess of 100 kpc from the galactic
plane, and its density in the plane is likely 
a fraction of a percent of the thin disk density 
(Norris 1999; Chen et al. 2001; Siegel et al. 2002; Larsen \&
Humphreys 2003; Robin et al. 2003; Cabrera-Lavers, Garz\'on 
\& Hammersley 2005). 

The derivation of stellar densities from star counts involves several
assumptions. For example, only very rough information on chemical composition
and luminosity is available from broad-band photometric observations alone,
forcing practitioners to adopt pre-determined relationships, and to rely on the
hypothesis that dwarfs outnumber evolved stars in any given field. 
Spectroscopic surveys of galactic stars, despite being of necessity biased
in order to sample low-density populations (such as the
galactic halo in the solar neighborhood), are extremely valuable to complement
star count analyses and unravel co-existing stellar populations.
Kinematics and chemical abundances expand our view of the Galaxy
from a snapshot provided by star counts to a more dynamical perspective,
allowing the study of not just the present structure of the Milky Way, but to
peer into its formation and evolution (see, e.g., Chiba \& Beers 2000, 2001;
Freeman \& Bland-Hawthorn 2002; Nordstr\"om et al. 2004).

Extensive spectroscopic surveys with high spectral resolution have so far been
primarily restricted to the solar neighborhood. Single-object spectrographs
delivering medium-resolution spectra have been the workhorse for studies of
large numbers of more distant stars (e.g., Beers, Preston, \& Shectman 1992;
Beers 1999; Wilhelm et al. 1999b; Brown et al. 2003; Christlieb 2003), but
recently, highly-multiplexed spectrographs 
have made possible an increase of the data
acquisition rate by several orders of magnitude. The near future is even more
exciting, with the advent of planned new instruments able to gather spectra for
thousands of targets simultaneously, over large fields, at high spectral
resolution, and without gaps in solid-angle (e.g., Hill \& MacQueen
2002; Moore, Gillingham, \& Saunders 2002). 

The now-completed Sloan Digital Sky Survey (York et al. 2000; SDSS) has imaged
about one fourth of the sky with five broad-band filters and obtained follow-up
intermediate dispersion (${\sf R} \equiv \lambda/{\rm FWHM} \sim 2000$)
spectroscopy of numerous targets. In its spectroscopic mode, up to 640 fibers
can be simultaneously positioned on the focal plane of the SDSS 2.5m telescope
to feed two identical spectrographs, both providing continuous coverage in the
range 381-910 nm (Newman et al. 2004). 

The criteria for selecting SDSS spectroscopic targets are rather complex (see
Strauss et al. 2002 and references therein). Galaxies and quasar candidates take
about 90\% of the fibers; the remaining fibers are used to observe the sky
background and galactic stars. The stars are either selected for being of
special interest (e.g., white dwarfs, blue horizontal-branch stars, carbon
stars, late-type M-dwarfs, brown dwarf candidates, etc.), or because they lie
within a narrow color range intended for refinement of reddening determinations
or flux calibration. In addition, a fraction of the SDSS quasar candidates
turned out to be stars (Stoughton et al. 2002). With exposure times of the order
of 45 minutes, the targeted stars have $V$ magnitudes in the range 14--22,
signal-to-noise ratios ($S/N$) between 4 and 60, and lie at distances of up to a
few hundred kiloparsecs from the galactic plane. More than $10^5$ stellar
spectra will be released to the public by the end of 2006. 
Nearly 50,000 stellar
spectra were made publicly available in the second data release (Abazajian et
al. 2004; hereafter DR2), and an additional $\sim$ 20,000 stellar spectra 
followed as part of DR3 (Abazajian et al. 2005). 
DR4 became public on July 2005
(Adelman-McCarthy et al. 2005), 
expanding the public database with over 12,000 new stellar spectra. 

Among the stellar targets in the SDSS database, low-mass stars are of particular
interest because they have long lives and thus 
can be used to trace the chemical
evolution of the Galaxy. In principle, the numbers of low-mass stars at
different metallicities reflects the history of star formation and galaxy
assembly. Accurate analysis of relative chemical abundances is now routinely
possible for F- and G-type stars within the framework of classical
one-dimensional model atmospheres (but see, e.g., Asplund 2005 for caveats).
Most of the stars in DR3 belong to these categories.

In this paper we begin a detailed exploration of this extensive database. 
We have used SDSS photometry and spectra to derive 
stellar atmospheric parameters,
radial velocities, and distance estimates for a significant fraction of the
stars included in DR3. The dataset and the analysis procedure are described in
\S \ref{method}. In \S \ref{comp}, we thoroughly test our derived parameters by
comparing with three spectroscopic libraries of nearby stars with 
reliable physical parameters. In \S \ref{dr3} we examine the distributions 
of stars in DR3, their iron abundances, and their kinematics at different 
distances from the galactic plane and from the galactic center. In \S
\ref{sum}, we conclude with a brief summary and comment on 
research to follow. Readers who
wish to skip the technical details might wish to resume in \S \ref{dr3}.

\section{Description of the dataset and our model-atmosphere 
	analysis}
\label{method}	

DR3 contains 528,640 spectra from 574 plugplates, covering 2527 deg$^2$ on
the sky. The SDSS pipeline classified 71,397 of the DR3 spectra as 
{\sc STAR}s. These data were pre-processed  and analyzed as described below. 

\subsection{Pre-processing of DR3 stellar spectra and $gr$ photometry}
\label{pre}

SDSS $ugriz$ photometry is available for all the objects with spectra. The
signal-to-noise ratio per pixel for the SDSS spectra is larger than 4 at
$g=20.2$, and the wavelength calibration is accurate to better than 5 km
s$^{-1}$. A number of improvements have been made to the SDSS spectroscopic
pipeline for DR2/DR3, in particular to the spectrophotometric calibration and to
the determination of radial velocities, as compared to data released in DR1 and
the early data release (EDR). For each plugplate, 16 fibers are allocated to
F-type stars that serve as reddening or spectrophotometric calibrators (in equal
proportions). The relative flux calibration is obtained by assigning a spectral
type to each calibration star, and using synthetic spectra based on Kurucz's
model atmospheres to remove the instrumental response. Absolute fluxes are
derived by forcing the $r$-band magnitudes computed from the corrected
calibration spectra to match the $r$ magnitudes from the calibrators measured by
the photometric pipeline. For more details we refer the reader to Stoughton et
al. (2002) and Abazajian et al. (2004, 2005).

Heliocentric velocity corrections have already been calculated and applied to
archival SDSS spectra, which are distributed on a logarithmic vacuum wavelength
scale. Doppler shifts in the stellar spectra are measured by cross-correlation
with observed templates in the SDSS pipeline. As part of our data
pre-processing, we change the wavelength scale to standard air (Edl\'en 1966),
and re-sample the spectra linearly.  Assuming that the SDSS spectrographs deliver
${\sf R} \simeq 2000$, independent of wavelength, we smooth the spectra to a
resolving power ${\sf R} \simeq 1000$ by convolution with a Gaussian profile
equivalent to ${\sf R}_{\rm G} = 1155$. This step is taken following tests with EDR
spectra, which indicated that for our purposes it is advantageous
to trade resolution for $S/N$. The uncertainties provided as part of
the SDSS spectra are also corrected for the change in resolution.
 
We first determine the velocity shifts of several strong features (H$\alpha$,
H$\beta$, H$\gamma$, H$\delta$, and CaII K) to obtain an independent measurement
of the stellar radial velocities and their uncertainties. The spectra in which
we could not identify Balmer lines 
were not considered for further analysis, as they
are either very hot, very noisy, exhibit emission features, or are too cool, and
thus unsuitable for the analysis techniques described below. The reference
wavelengths were derived from the solar spectrum observed at much higher
resolving power and smoothed to the appropriate resolution. The spectra are then
shifted in velocity to the rest frame and re-sampled with three data points per
resolution element. 

\begin{figure}[t!]
\includegraphics[angle=90,width=8cm]{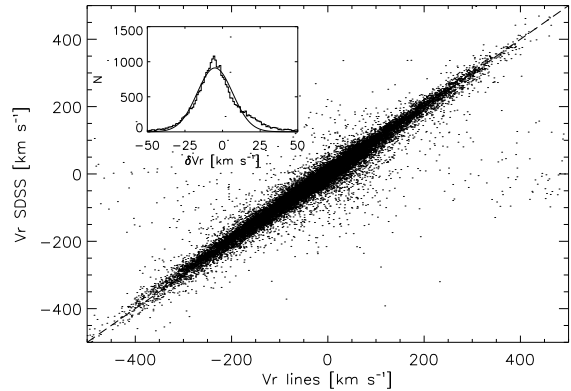}
\caption{Radial velocities derived from the Doppler shift of strong lines
and from the SDSS pipeline. A small 
systematic zero-point offset is present (see text). 
The dashed line has a slope of
one. The inset shows an histogram of the velocity differences 
and a Gaussian fit. \label{vr}}
\end{figure}

We identified Balmer lines and derived radial velocities for 44,175 DR3 stars.
Fig. \ref{vr} compares our derived velocities with those provided by the SDSS
pipeline ($zc$, as they are coded as a redshift in the FITS headers) for 30,589
objects with colors in the range $0.0 \lesssim (g-r) \lesssim 0.7 $ mag. This
corresponds, roughly, to stars with $5000 < T_{\rm eff} < 8000$ K.
On average, our radial velocities are smaller by roughly 5 km s$^{-1}$. Removing
the zero-point offset, and excluding far outliers, 
the two scales agree with a
1$\sigma$ scatter of 12 km s$^{-1}$, although the residuals are not Gaussian, as
illustrated in Fig. \ref{vr}. Most of the stars are contained within the range
$-500 < V_r < 500$ km s$^{-1}$, but there are about 20 objects with velocities
between $500 < V_r < 1000$ km s$^{-1}$ for which there is good agreement between
the values from the pipeline and our measurements. Such velocities exceed the
escape velocity from the Galaxy, and therefore a few of these objects may belong
to the same class as the hyper-velocity star recently reported by Brown et al.
(2005). The majority appear to be galaxies in the Virgo
cluster, mistaken for stars by the SDSS classifiers.

Most of the spectral information that can be used to constrain the atmospheric
parameters resides in the blue region, but the $S/N$ peaks in the red for
late-type stars. As a compromise, we trimmed the spectra to the region 440-550
nm, increasing the analysis speed by a factor of about five as compared to an
analysis of the full spectral range. The spectra are normalized by iteratively
fitting a sixth-order polynomial to the pseudo-continuum, clipping off points
that lie farther than 1$\sigma$ below, or 4$\sigma$ above the fitting curve. As
the absolute flux calibration is tied to the photometric scale of the imaging
survey, and the relative spectrophotometry is based on fluxes calculated in a
similar way as those we match to the observations in our analysis, little
information is lost by applying this correction. In addition, the impact of
reddening on the spectra is essentially removed. 

In addition to the spectra, we also use SDSS $gr$ photometry to constrain the
stellar parameters. Restricting the photometry to the $g-r$ color index makes it
possible to perform a fair comparison with the libraries of nearby stars
described below, which in most cases only include $BV$ photometry. The reddening
in $B-V$ was interpolated from the maps of Schlegel, Finkbeiner \& Davis (1998),
with the following transformations for the SDSS $g$ and $r$ passbands

 \begin{equation}
\begin{tabular}{lll}
 $A_g$ &$=$& $3.793 ~E(B-V)$\\
 $A_r$  &$=$& $2.751 ~E(B-V).$\\
\end{tabular}
 \end{equation}

\noindent Given that the location of the vast majority of our targets
are at distances larger than 0.5 kpc, and the relatively high galactic latitude
of the fields, it seems appropriate to apply the full extinction from the maps. 

The pre-processing of DR3 spectra described above can be summarized
as follows: we smooth, normalize, velocity correct, re-sample, and truncate 
the  spectra to the  spectral region 440--550 nm. 
Point Spread Function $gr$ photometry is also included in the analysis,  
after correcting for interstellar extinction. Throughout the paper, we
refer exclusively to photometry corrected for reddening.

\begin{figure}[t!]
\includegraphics[angle=0,width=8cm]{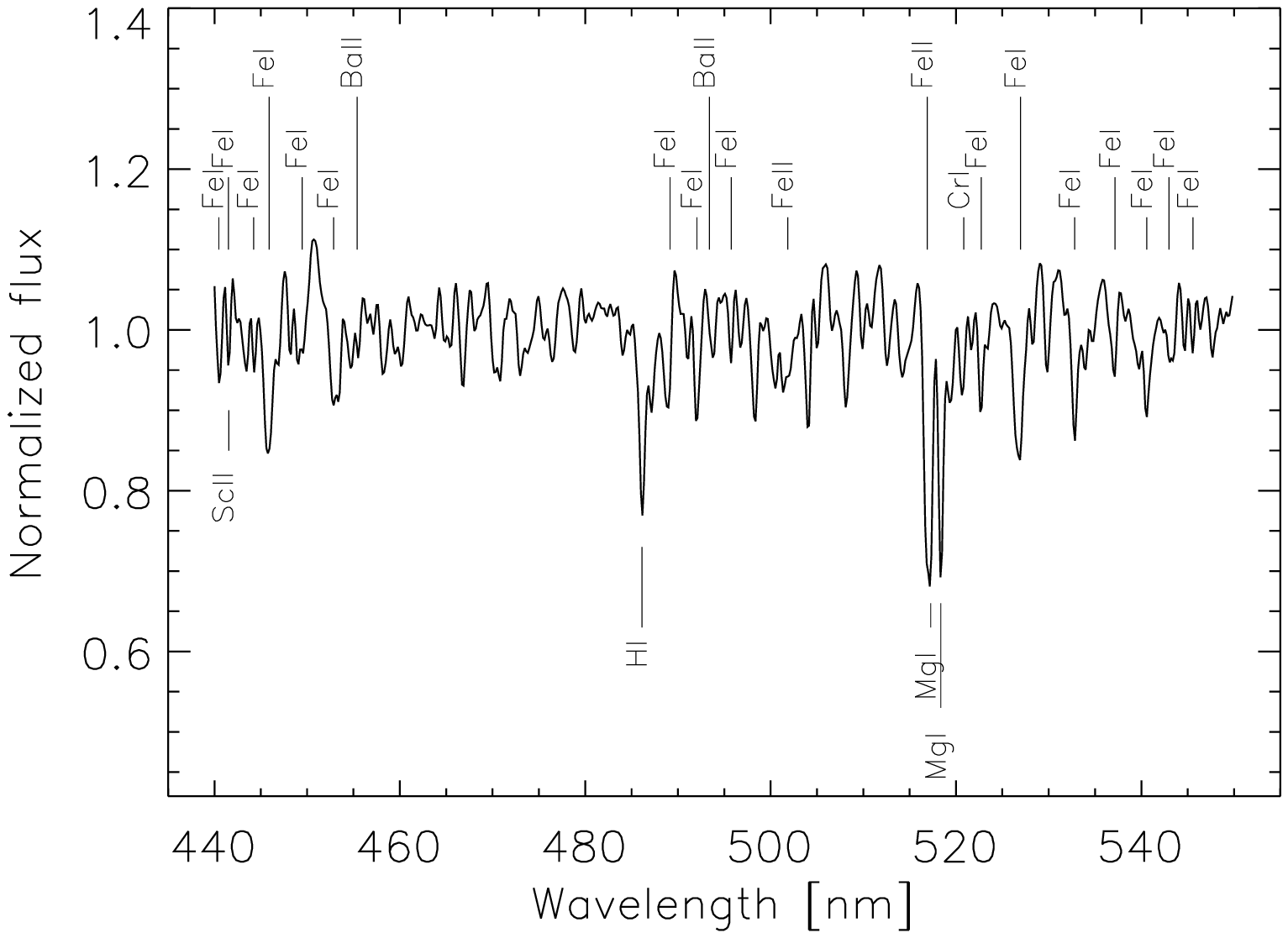}
\includegraphics[angle=0,width=8cm]{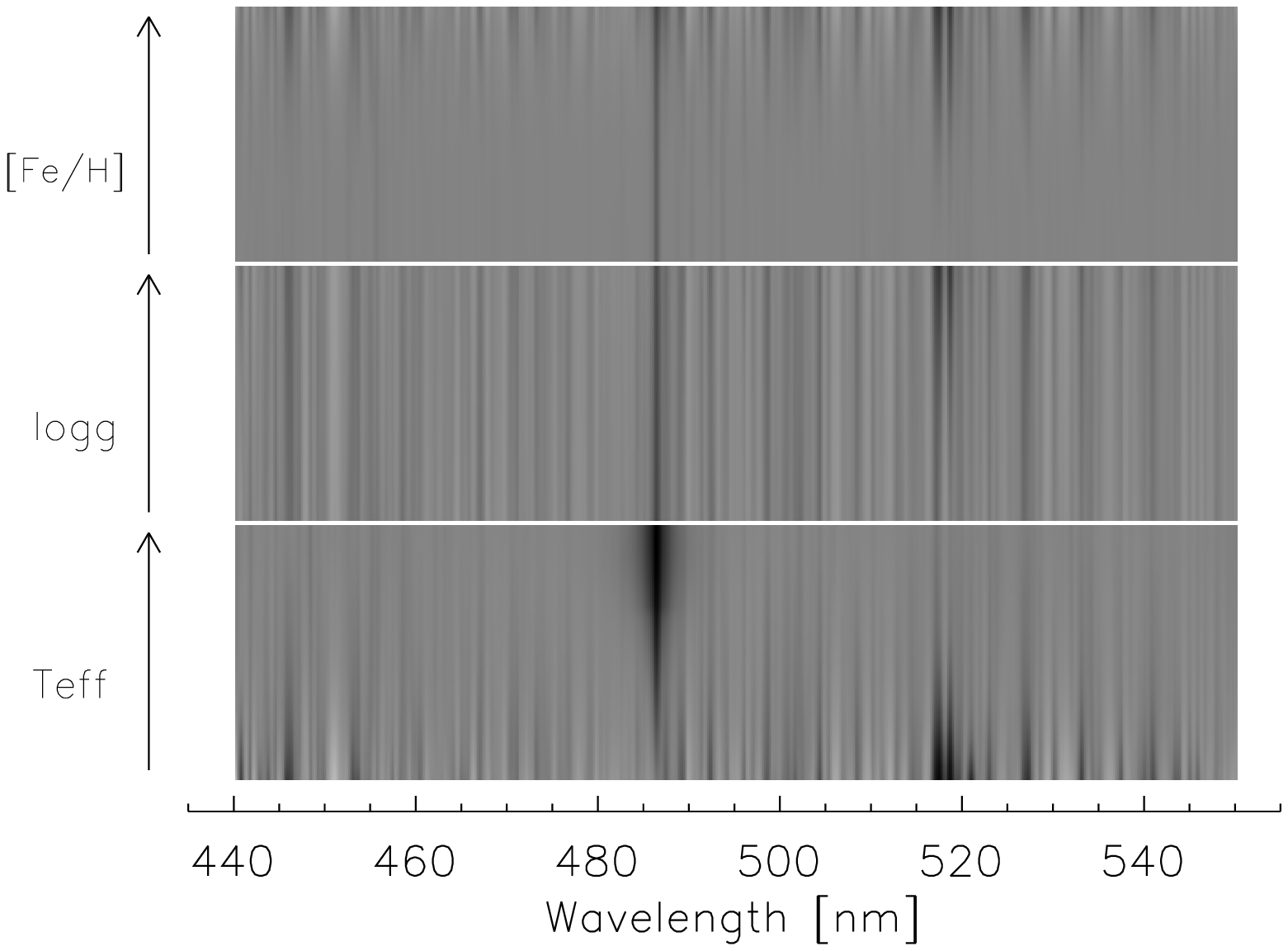}
\caption{{\it Upper panel} Computed spectrum for a solar-like star; the
strongest features due to single transitions are identified. {\it Lower panel}
Variation of the continuum-corrected flux as a function of each of the
three atmospheric parameters considered here, with the others fixed at
near-solar values. \label{id}}
\end{figure}

\subsection{Analysis}
\label{ana}

Our analysis combines spectra and photometric indices from SDSS. We will refer
to the data vector as ${\bf T} \equiv \{ (g-r),S_1,S_2,...,S_m\}$, where the
$S_i$ are the continuum-corrected spectral fluxes, and $m=667$ for our choice of
resolving power and sampling. We model {\bf T} with plane-parallel
line-blanketed model atmospheres in local thermodynamical equilibrium (LTE) as a
function of the stellar parameters {$T_{\rm eff}, \log g$, and
[Fe/H]}\footnote{Following standard use, $T_{\rm eff}$ is the effective
temperature, $g$ is the surface gravity (given in cm s$^{-2}$ throughout the
text), and [Fe/H] $= \log_{10} ({\rm N(Fe)/N(H)}) - \log_{10} ({\rm N(Fe)/N(H)})
_{\odot}$, where N represents number density.}. The model atmospheres are
extracted from the Kurucz (1993) grid, which was calculated adopting a
mixing-length $l/H_p = 1.25$, and a micro-turbulence of 2 km s$^{-1}$. A
parallel grid of low-resolution (${\sf R} \sim 300$) spectra, computed with
detailed line and continuous metal-opacities, has been provided by Kurucz (1993).
Synthetic photometric indices are derived from these spectra and the SDSS
filter responses (Strauss \& Gunn 2001\footnote{\tt http:
//archive.stsci.edu/sdss/documents/response.dat}). These indices differ from
those presented by Lenz et al. (1998), as the detailed filter responses for the
2.5m telescope were not available at that time. The atmospheric structures are
used to produce a second grid of LTE synthetic spectra with a resolving power
(${\sf R} = 1000$) and spectral coverage (440--550 nm) matching the
pre-processed SDSS spectra. To produce this second grid, we employ the code {\tt
synspec} (Hubeny \& Lanz 2000), and very simple continuous opacities: H,
H$^{-}$, H Rayleigh and electron scattering (as described by Hubeny 1988).
Atomic line opacity is considered with 37,566 transitions, neglecting molecular
opacity, treating Balmer line profiles as in Hubeny, Hummer \& Lanz (1994).

The grids of calculated spectra considered here span the ranges [4500,9250] ~K
in $T_{\rm eff}$, [1.5,5.0] in $\log g$, and [$-4.83$,$+0.67$] in [Fe/H]. The
upper panel of Fig. \ref{id} shows a calculated spectrum for a solar-like star
after correcting the continuum shape. This correction is applied to the absolute
fluxes in exactly the same manner as to the observed (flux-calibrated) spectra
(see \S \ref{pre}) to ensure consistency. The strongest features due to single
transitions have been identified in the figure. Most of the lines are due to
iron, and many of the observed features result from multiple overlapping
transitions. The lower panel shows three images created by stacking the
continuum-corrected spectra (with a grey-scale indicating the flux level) as a
function of each of the three atmospheric parameters considered, with the other
two fixed at solar values. It is immediately apparent that variations in surface
gravity produce the most subtle changes in the spectrum, and that the presence
of the Mg I b triplet is key to our ability of constraining this parameter from
the observations. The wings of the Mg I b lines are collisionally enhanced, and
dominate the gravity-driven changes on the strength of the lines (e.g., Fuhrmann
et al. 1997).

Model spectra and photometry for sets of parameters off the grid nodes are
derived by interpolation (Allende Prieto 2004). The  high speed required
to classify such a large sample is achieved at the expense of using a fairly
coarse mesh, with $12\times 15\times 19$ nodes in $T_{\rm eff}-\log g-$[Fe/H],
selected to match the nodes of the available grid of model atmospheres,
avoiding interpolations in the atmospheric structures. 

We adopted reference solar abundances as in Asplund, Grevesse \& Sauval (2004).
These solar abundances include the recently revised values of the photospheric
C, O, and Fe abundances, which are roughly 0.2 dex lower than those used by
Kurucz (1993). Thus, we corrected for this amount in the selection of the
atmospheric structures and low-resolution fluxes from Kurucz's grid. Some
chemical elements have abundance ratios to iron that are non-solar in metal-poor
stars; this is largely ignored in our modeling. However, we do take into account
enhancements to the abundances of the $\alpha$ elements Mg, Si, Ca, and Ti in
metal-poor stars when calculating synthetic spectra. Following Beers et al.
(1999), we adopt  

\begin{equation}
[\alpha/{\rm Fe}] = \left\{ 
\begin{tabular}{cc}
  		           0 		& if [Fe/H] $\ge  0 $  	\\
  		        		      			& \\
        	      $-0.267$ [Fe/H] 	& if $-1.5 \le$ [Fe/H] $< 0$ \\
 	  		      			& \\
  		            $+0.4$ 	& if  [Fe/H] $< -1.5$.   \\
\end{tabular}
\right.
\label{law}
\end{equation}

\noindent The above recipe will work well for thin-disk stars and 
low-metallicity halo stars, but it will underestimate the [$\alpha$/Fe] ratios
in halo and thick-disk stars with [Fe/H] $>-1$. The systematic errors will
affect thick-disk stars the most, given that they concentrate in that
metallicity regime. The [Mg/Fe] abundance ratio for thick-disk members remains
as high as $\sim 0.3$ dex up to [Fe/H] $\simeq -0.4$ (Fuhrmann 1998; Prochaska et
al. 2000; Bensby, Feltzing \& Lundstr{\" o}m 2003; Reddy et al. 2003;
Brewer \& Carney 2005; Reddy, Lambert \& Allende Prieto 2005), and Eq.
\ref{law} would therefore underestimate this ratio by $\sim 0.2$ dex for such
stars. Because the strength of the wings of the Mg I b lines is proportional to
both the Mg abundance and the density of the collisional perturbers (mostly H
atoms), underestimating the former would be compensated by a similar error of
opposite sign in the gas pressure, resulting in surface gravity estimates which
are too high by about 0.2 dex. Although such biases are not negligible, 
they are smaller than the typical uncertanties in the surface gravities 
that we derive from the SDSS data.

We  search for the  model parameters that minimize the 
distance between the model flux vector {\bf T} ($T_{\rm eff}, \log g$, [Fe/H]),
and the  observation vector {\bf O} in a  $\chi^2$ fashion
\begin{equation}
\mu =  \sum_{i=1}^{m+1}{\frac{W_i}{\sigma_i^2} (O_i - T_i)^2} 
\label{fit}
\end{equation}
 
\noindent where the weights $W_i$ were set equal for 
all the data points in a spectrum, 
about two orders of magnitude larger for the photometric 
index $(g-r)$, and normalized $\sum W_i = m+1$.

\begin{figure}
\includegraphics[angle=0,width=8cm]{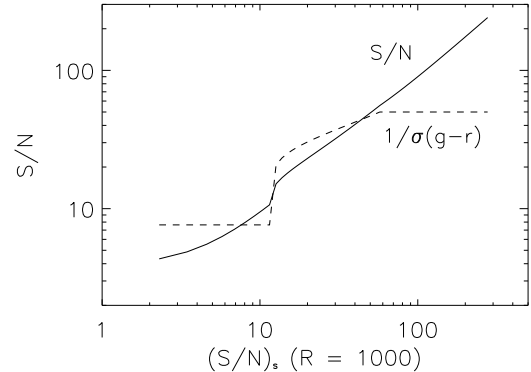}
\caption{The weighted $S/N$ for DR3 stars as used here 
is defined as the square root of a linear
combination of the signal-to-noise of the spectrum $(S/N)s$ and that of the
color index $(g-r)$, which is tied to the former by an empirical
recipe (Eq. \ref{snrp}).  \label{snr}}
\end{figure}

The search is accomplished using the Nelder-Mead simplex method
(Nelder \& Mead 1965). This optimization algorithm, combined
with interpolation, performs well in terms of accuracy and speed, processing
several stars per second on a modern workstation.
Error bars for the spectral fluxes are available as part of
the SDSS data, and we have accounted for the reduction of resolution
in our pre-processing (see \S \ref{pre}). The weighted $S/N$ ratio is derived 
\begin{equation}
 S/N =  \sqrt{\frac{1}{m+1} \sum_{i=1}^{m+1} \frac{W_i}{\sigma_i^2}},
\label{snrdef}  
\end{equation}
\noindent and $\sigma_1 \equiv \sigma_{(g-r)}$ is estimated
 from the signal-to-noise ratio of the spectrum 
\begin{equation}
 (S/N)_s =  \sqrt{\frac{1}{m} \sum_{i=2}^{m+1} \frac{1}{\sigma_i^2}},
\label{snr2}  
\end{equation}
\noindent and an empirical relationship for SDSS data
\begin{equation}
\sigma_{(g-r)} =  \left\{ 
\begin{tabular}{cc}
    0.131 		& if 	$(S/N)_s$ $< 11.500024$  	\\
		      	& 	 \\
    0.02 		&  if   $(S/N)_s$ $> 57.5$   \\
  			& 		 \\ 	  		         
${\displaystyle   0.02 - {\frac{1}{130.4}}}$ & ${\displaystyle 
 \ln  \left( \frac{(S/N)_s - 11.5}{46} \right)}$  else.  \\
\end{tabular}
\right.
\label{snrp}
\end{equation}
\noindent The $S/N$ of the spectra
smoothed to ${\sf R} \simeq 1000$ is approximately 2.3 times higher
than that of the original data. Fig. \ref{snr} illustrates the
weighted $S/N$ of our processed spectra as a function of $(S/N)_s$. 
The vast majority of the stars  have $20 <S/N < 100$;  stars with $S/N<30$ 
are not considered in this study.
 
\begin{figure*}
\begin{tabular}{cc}
\includegraphics[angle=0,width=8cm]{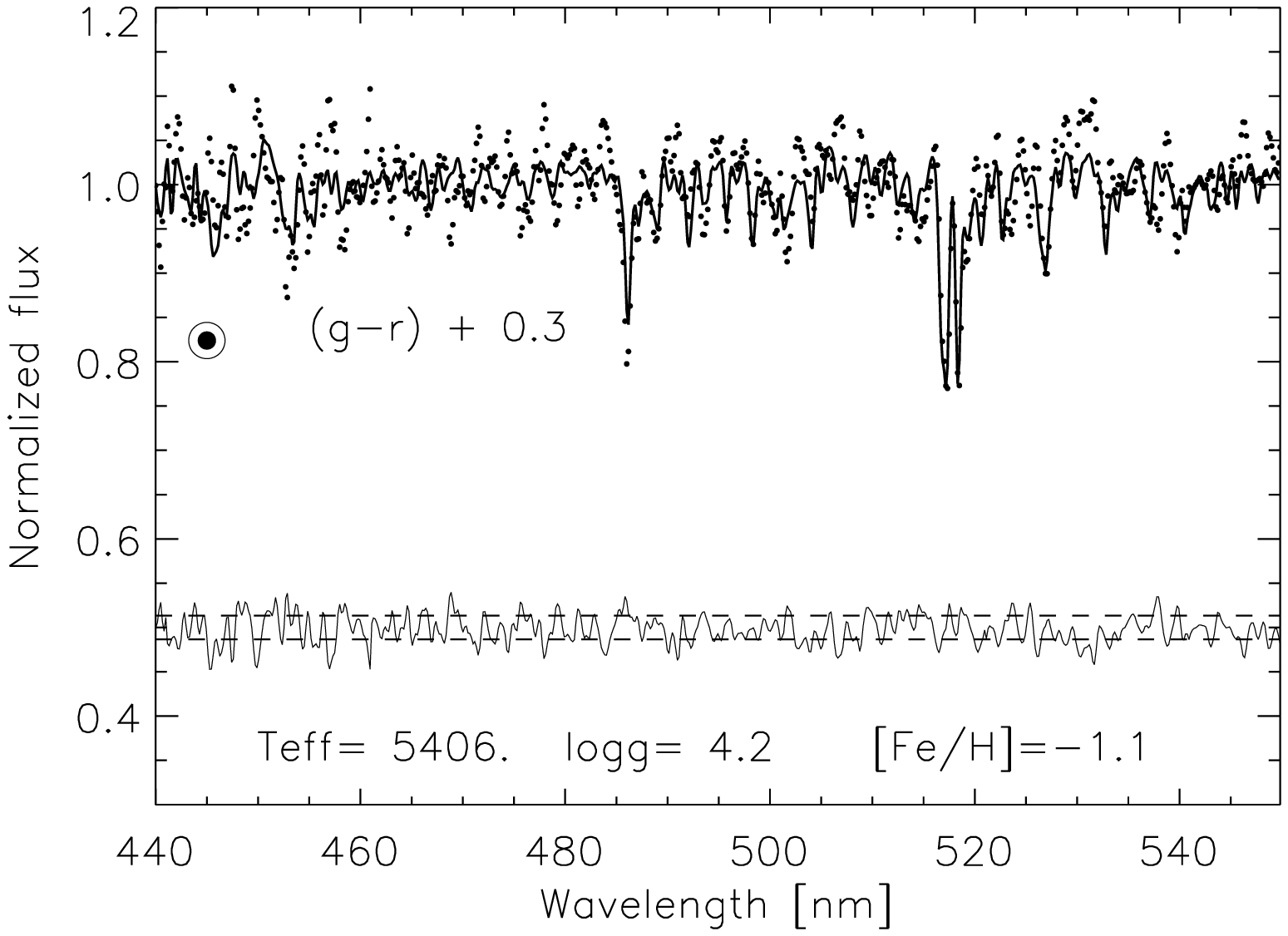} & 
\includegraphics[angle=0,width=8cm]{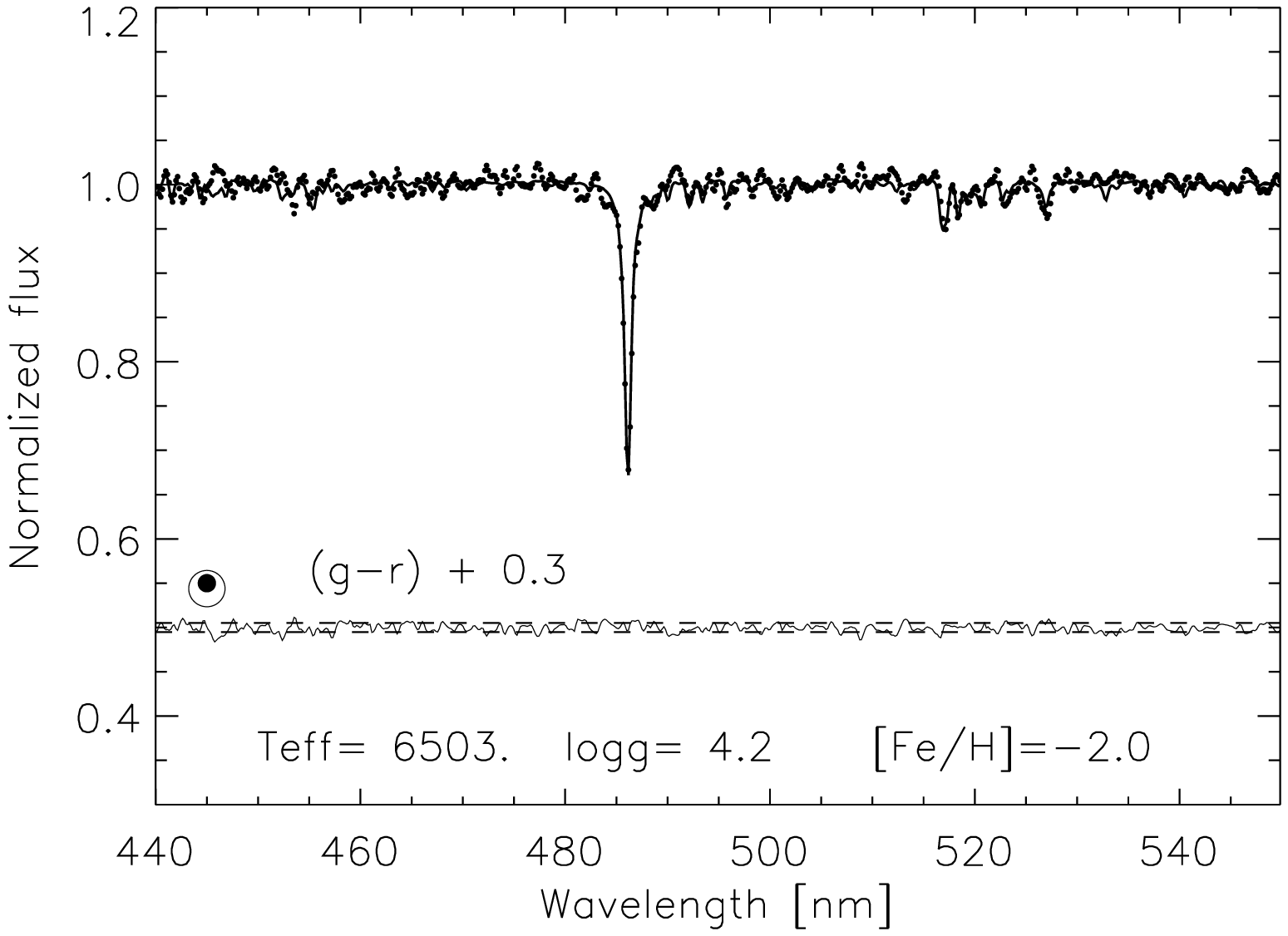} \\
\includegraphics[angle=0,width=7cm]{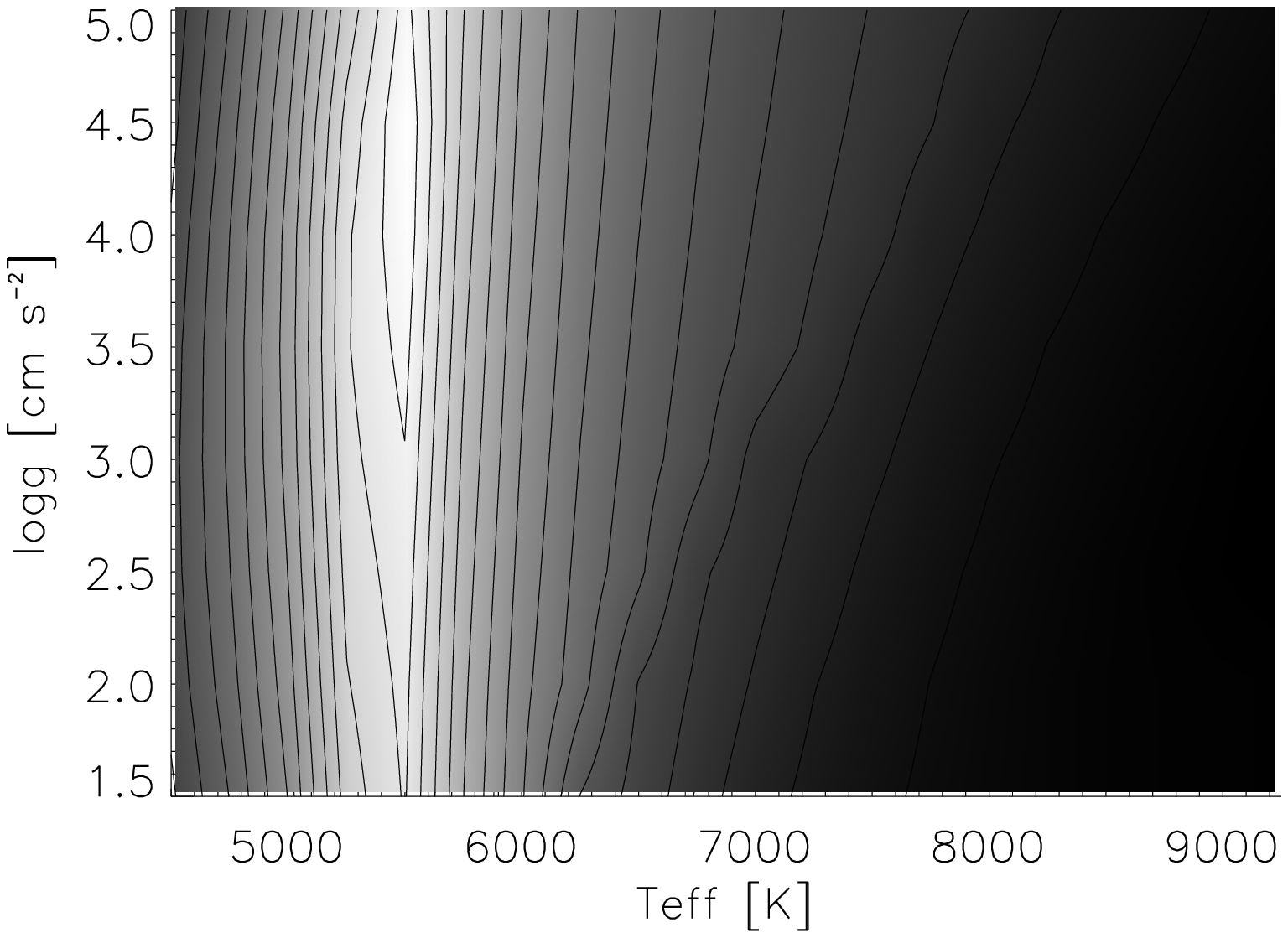}  &
\includegraphics[angle=0,width=7cm]{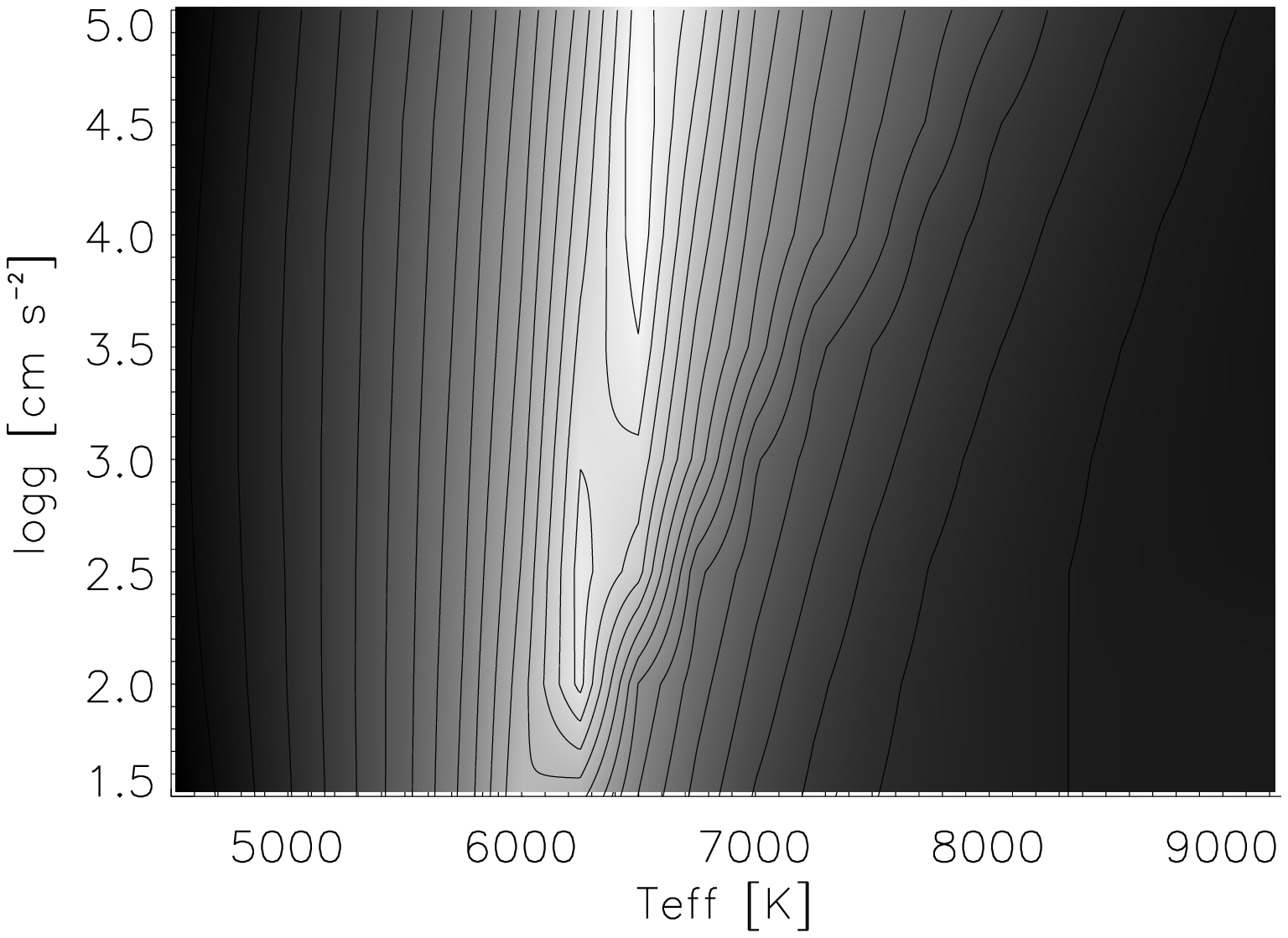}  \\
\includegraphics[angle=0,width=7cm]{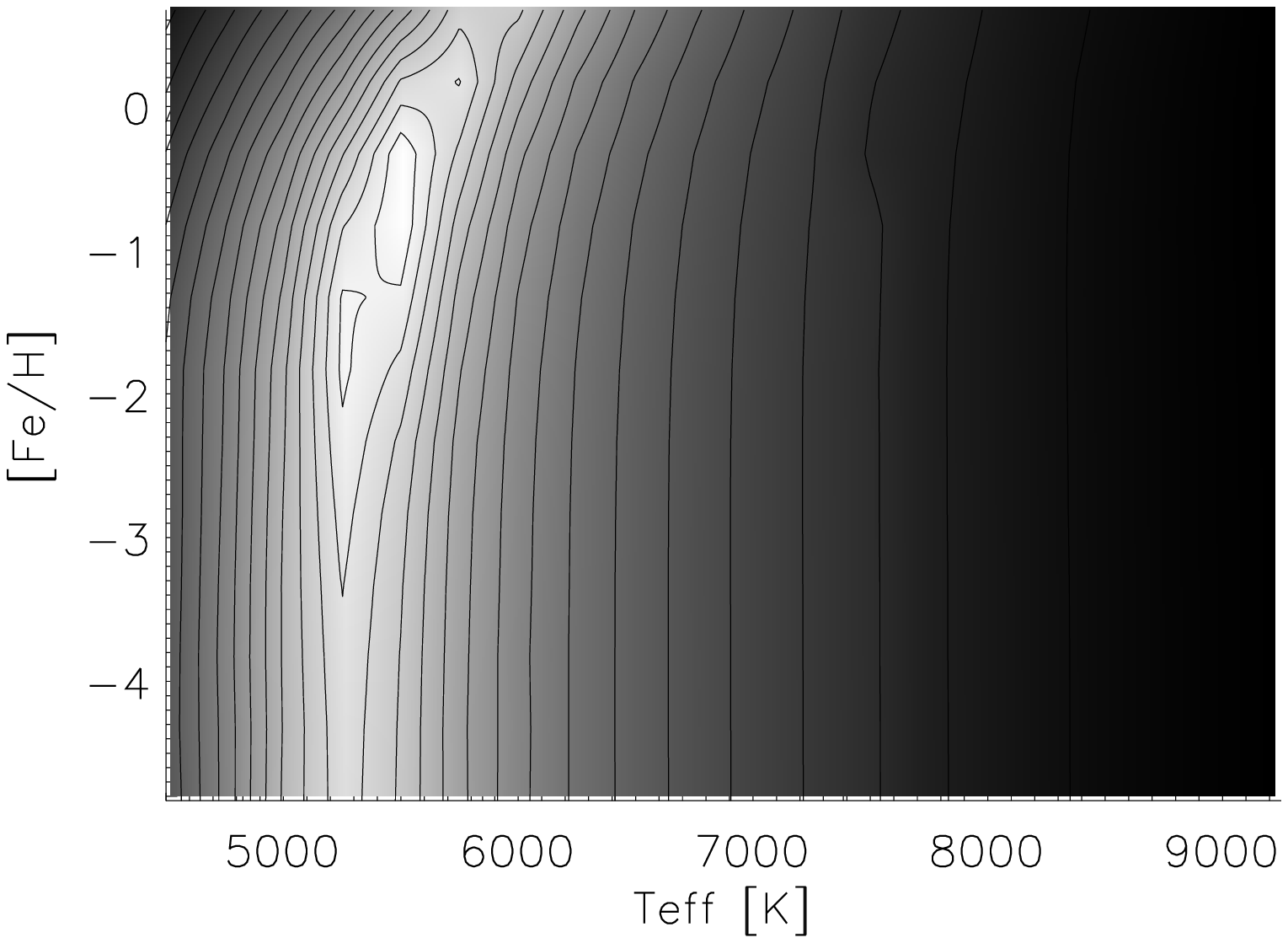} & 
\includegraphics[angle=0,width=7cm]{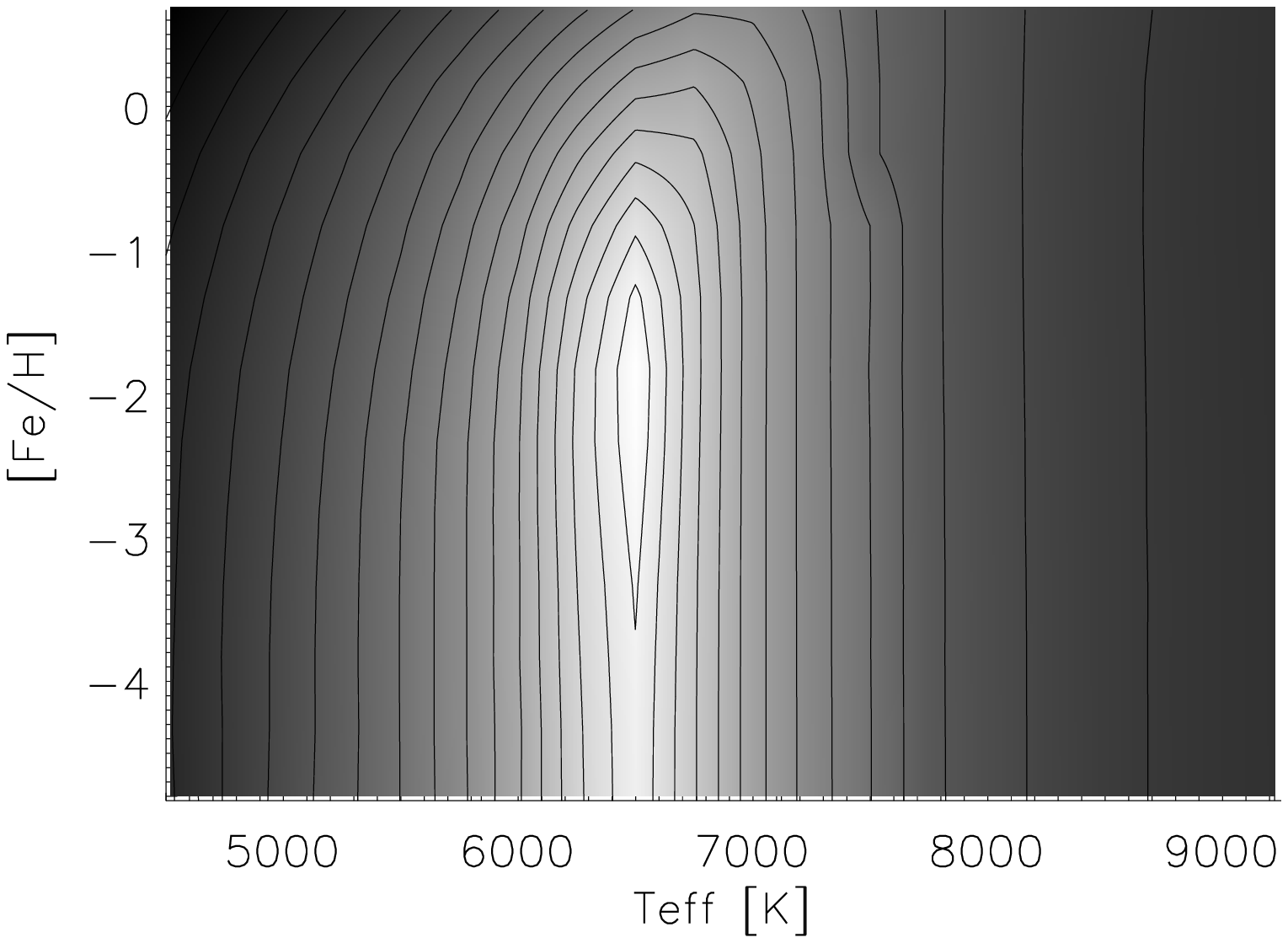}  \\
\includegraphics[angle=0,width=7cm]{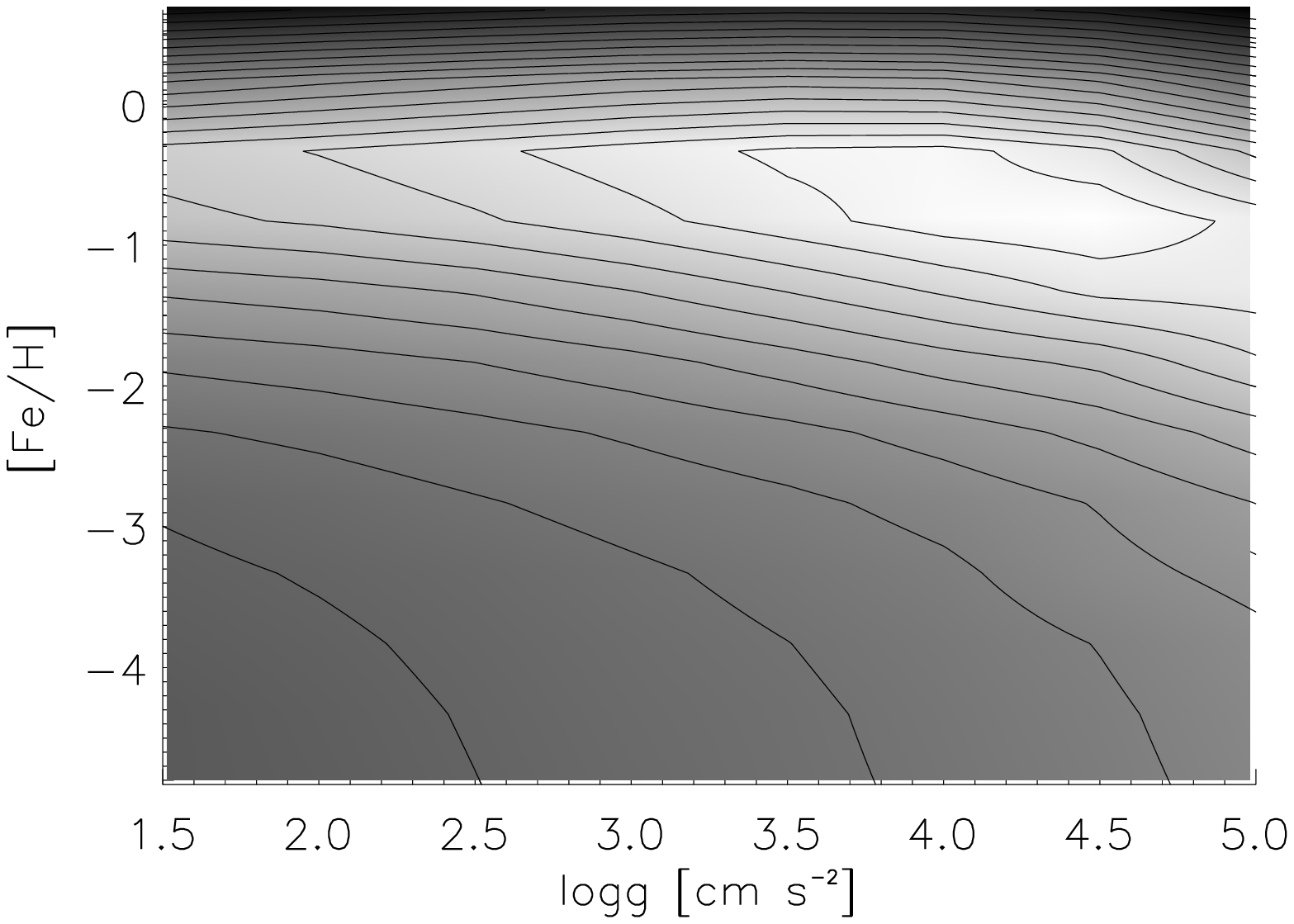}  & 
\includegraphics[angle=0,width=7cm]{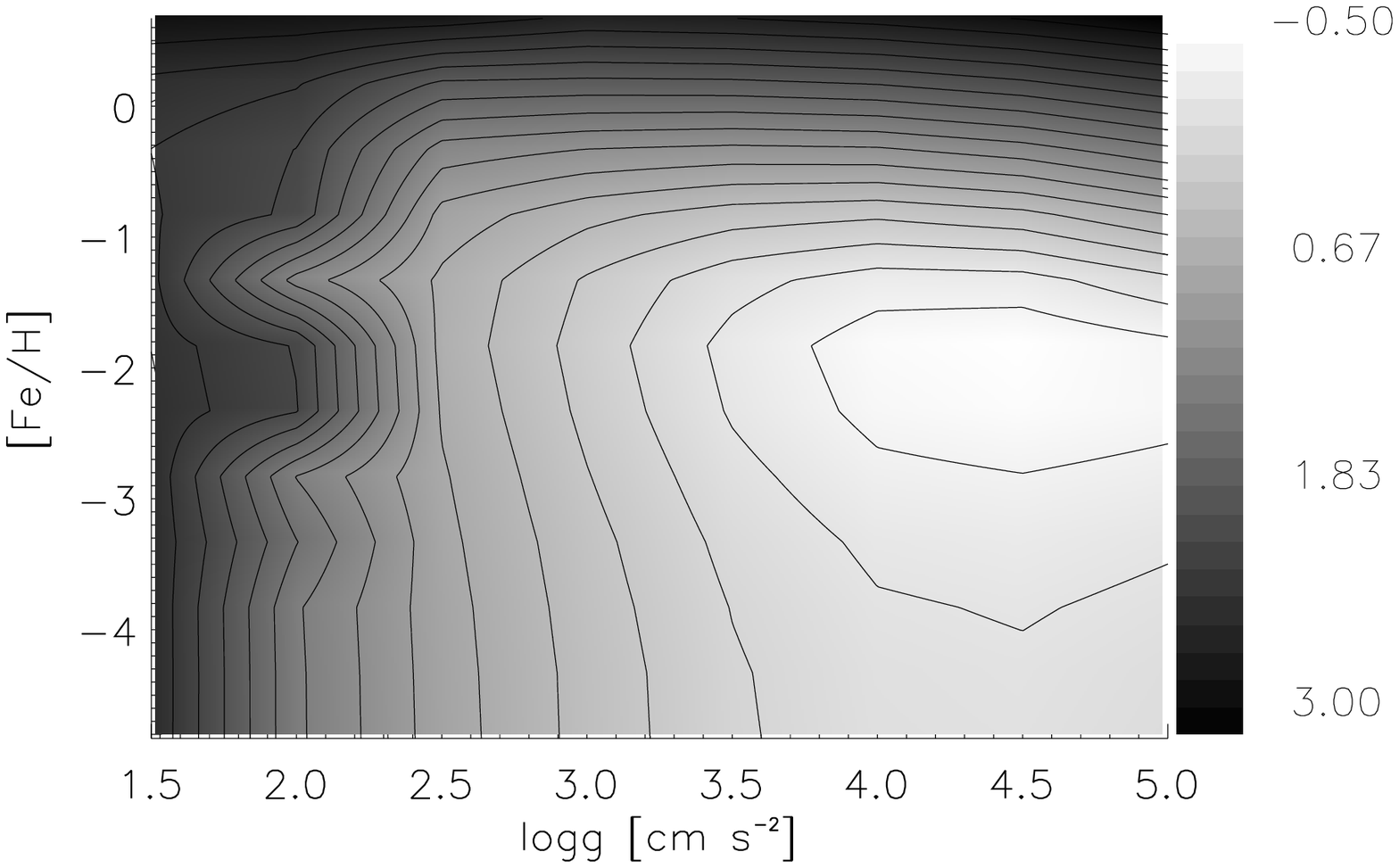}  \\
\end{tabular}
\caption{Two examples of our data analysis. 
The graphs on the left correspond to  the object
SDSS J135701$+$520018.4, 
a G-type subgiant with a metallicity [Fe/H] $\sim -1$ ($T_{\rm eff}=5406$ K,
$\log g = 4.2$, [Fe/H] $=-1.1$); the right-hand-side panels are for
SDSS J135526$+$512938.1, 
an F-type metal-poor dwarf 
($T_{\rm eff}=6503$ K, $\log g = 4.2$, [Fe/H] $=-2.0$). The 
top panels compare the observed (filled circles) and the best-fitting spectra 
(solid line), as well as the observed $g-r$  (large filled circles) with 
the best-fitting indices (open circles), shown at an arbitrary 
wavelength. The lower solid
lines in each of the top panels indicates the residuals obtained from the
best-fitting spectra plus 0.45, while the dashed lines mark the average error
of the fluxes. The lower panels
show the variations in the logarithm of the reduced $\chi^2 = \log (\mu/m)$ in 
planes that cross the optimal
solutions. Three-sigma errors approximately correspond to 
deviations of $+1$ from
the minimum in the gray scale. \label{fits}}
\end{figure*}

The effective temperature, $T_{\rm eff}$, is fairly well-constrained for most
stars. The metallicity, [Fe/H], is tightly confined for the metal-rich and
cooler stars in our sample, but degrades for warmer stars, and especially so for
the most metal-poor cases. The surface gravity, $\log g$, is the most poorly
constrained parameter. Fig. \ref{fits} shows two examples of the matching
between observed and model fluxes, as well as the variations of $\log \chi^2$ in
planes across the optimal solutions. The observed and calculated photometric
indices, $(g-r)$, are shown at an arbitrary wavelength in the upper panels.
Other techniques for obtaining estimates of atmospheric parameters have also
been explored, and we compare them with our minimum-distance method in Section
\ref{other}.

\begin{figure}
\includegraphics[angle=0,width=8cm]{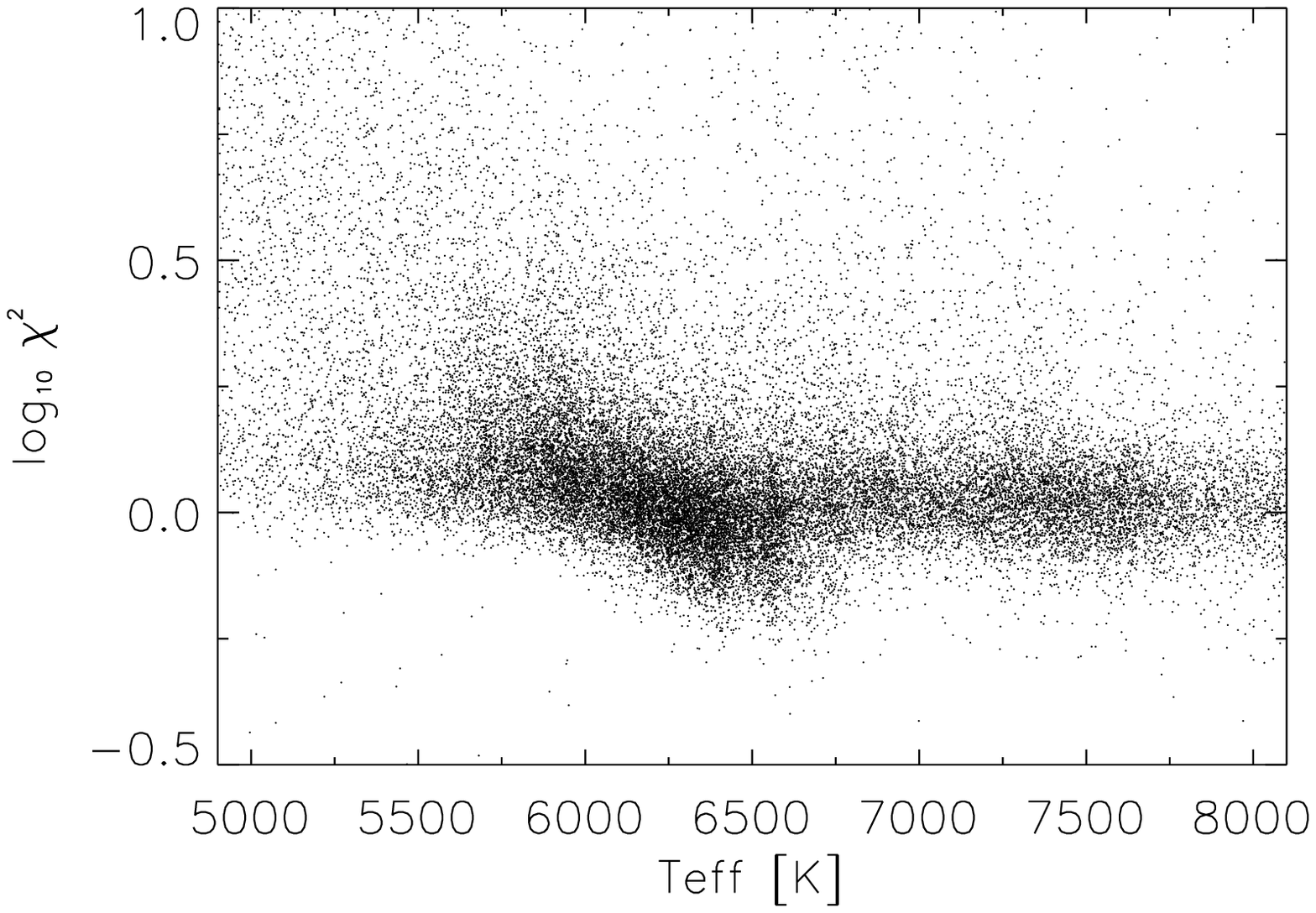}
\caption{Variation of the reduced $\chi^2$ as a function of the
effective temperature for the sub-sample of 44,175 DR3 stars with spectra 
for which we detected Balmer lines.  \label{q}}
\end{figure}

Fig. \ref{q} shows the final reduced $\chi^2 = \mu/m$ for the sample. Most stars
have late-F and early-G spectral types, and few are warmer than about 7000 K.
Late-G and early-K stars are fit somewhat worse than warmer spectral types,
mostly due to an increased importance of the metal line opacity, which is
afflicted by poor and missing atomic data. The two cases illustrated in
Fig. \ref{fits} have a reduced $\chi^2$ of $1.2$ and 0.7, for the more
metal-rich and more metal-poor star, respectively.

Once the atmospheric parameters are defined, we make use of stellar evolution
calculations by the Padova group (Alongi et al. 1993; Bressan et al. 1993;
Bertelli et al. 1994; Fagotto et al. 1994) to find the best estimates for other
stellar parameters, such as radius, $M_V$, mass, etc. With the atmospheric
parameters and their uncertainties in hand we define a normalized probability
density distribution that is Normal for $T_{\rm eff}$ and $\log g$, and a boxcar
function in $\log (Z/Z_{\odot})$ 
 
 \begin{eqnarray} 
P   & \displaystyle \propto
\exp \left[ -\left(\frac{T_{\rm eff}-T_{\rm eff}^{*}}{\sqrt{2} \sigma(T_{\rm eff})}\right)^2\right]
	\\
    & 
\displaystyle \times \exp \left[ -\left(\frac{\log g - \log g^{*}}{\sqrt{2} \sigma(\log g)}\right)^2\right]
 B\left(\log (Z/Z_{\odot})\right),
\nonumber 	 	 
\label{triplegauss}
\end{eqnarray}
 
\noindent  which is then used to find the best estimate 
of a stellar parameter $X$ by integration over the space ($Z/Z_{\odot}$, Age,
and initial mass $M$) that characterizes the stellar isochrones of Bertelli et
al. (1994):
 
\begin{equation}
\bar{X} = 
\int\int\int X P(Z/Z_{\odot},Age,M) d(Z/Z_{\odot}) d(Age) dM.
\end{equation} 
 
Our method follows that described in the Appendix A of Allende Prieto, Barklem,
Lambert, \& Cunha (2004a), where more details can be found; in our case we
adopted $\sigma(T_{\rm eff})= 150$ K, and $\sigma(\log g)=0.3$ dex for all
stars. The boxcar is a justified simplification, and we adopted a width of 0.3
dex. The isochrones employed do not consider enhancements in the abundances of
the $\alpha$ elements for metal-poor stars. To compensate for this effect we
follow Degl'Innocenti, Prada Moroni, \& Ricci (2005) and equate

\begin{equation}
Z/Z_{\odot} = 10^{\rm [Fe/H]} (0.659 \times 10^{\rm [\alpha/Fe]} + 0.341),
\label{enhance}
\end{equation}

\noindent with [$\alpha$/Fe] as given in Eq. \ref{law}.
The isochrones have metallicities between $0.02 \le Z/Z_{\odot} \le 2.5$. For
stars with metallicities  [Fe/H] $< -2$, we use isochrones with
$Z/Z_{\odot}=0.02$. We do not expect significant changes in the structure and
evolution of low-mass stars between this metallicity levels and zero
metallicity, hence this approximation is unlikely to be a source of large
systematic errors. Our tests indicate that use of updated isochrones with
similar abundance ratios published by the Padova group (e.g., Girardi et al.
2004) would lead to very similar results.

From the magnitudes in the SDSS passbands we estimate the Johnson $V$ magnitudes
of the stars using the transformation derived by Zhao \& Newberg (private
communication; see Eq. \ref{zn2} below). This transformation, and others which
are part of the same work, were derived from a sample of 58 metal-poor and field
horizontal branch stars previously identified in the HK survey (Beers et al.
1999) with both $UBV$ and $u^{*}g^{*}r^{*}i^{*}z^{*}$ photometry. They should be
considered as approximate for our purposes, given that they are not referred to
the exact same photometric system as the DR3 magnitudes
($u^{*}g^{*}r^{*}i^{*}z^{*}$ vs. $ugriz$). Knowing $M_V$ and the reddening, it
is then straightforward to derive distances.
 

\section{Reference calibrations and consistency checks}
\label{comp}

We have selected three libraries of stellar spectra for checking our analysis
procedure: the Indo-US library of Coud\'e feed stellar spectra (Valdes et al.
2004, hereafter {\tt Cflib}), the Elodie library (Prugniel \& Soubiran 2001,
hereafter {\sc elodie}), and the Spectroscopic Survey of Stars in the Solar
Neighborhood (Allende Prieto et al. 2004, hereafter S$^4$N). These libraries
include photometry; in particular, all of them list Johnson $B$ and $V$
magnitudes, which can be used to estimate $(g-r)$, with the transformation
derived by Zhao \& Newberg:
\begin{equation}
	(g-r) = 1.043(B-V)-0.185.
\label{zn}	
\end{equation}

In addition to spectra and photometry, the chosen libraries provide catalogs of
atmospheric stellar parameters that can be considered reliable, based on a
consistent analysis of the spectra in the library and/or collated from
high-dispersion studies in the literature. Our comparison with the libraries of
nearby stars, and our later analysis of stellar spectra in DR3, is restricted to
the range $5000 < T_{\rm eff} < 8000 $ K. Outside these limits in effective
temperature, our results degrade significantly, due to systematic differences in
the temperature scale on the warm side, and a degeneracy between metallicity and
temperature for the cool stars.  Work is underway to improve these limitations
for future analysis of SDSS stars.

\subsection{{\tt Cflib}}

\begin{figure}[t!]
\begin{center}
\includegraphics[angle=0,width=8cm]{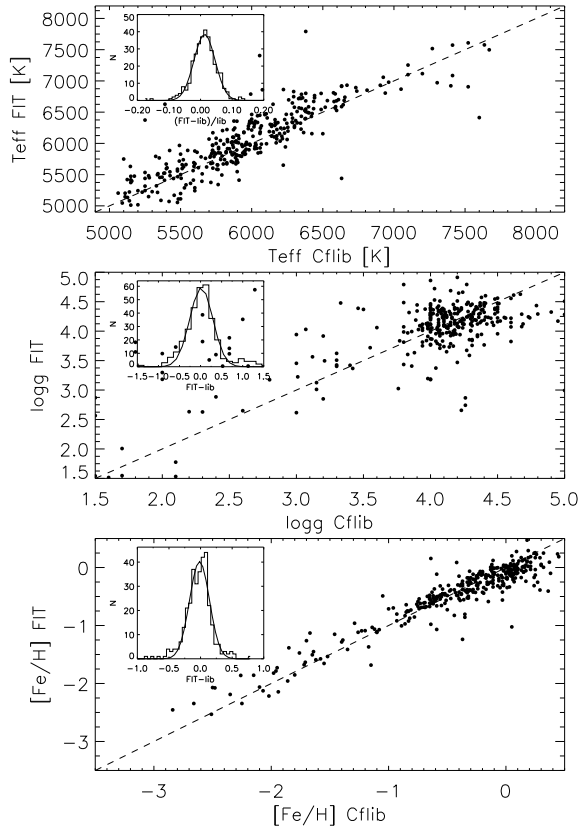}
\caption{Comparison between our derived 
metal abundances (FIT) and those from the catalog of the Indo-US 
library of Coud\'e feed 
stellar spectra ({\tt Cflib}). 
The dashed line has a slope of
unity. The inset graph shows the abundance differences, 
and a least-squares fitting to a Gaussian curve. \label{cflib}}
\end{center}
\end{figure}

{\tt Cflib} contains 885 stars with a spectral coverage similar to the SDSS
spectra, gathered with the 0.9m Coud\'e feed telescope at Kitt Peak National
Observatory
\footnote{\tt http://www.noao.edu/cflib/}. 
The $S/N$ ratio is relatively high, 
typically larger than 200, and the spectral
resolution is about 1.2 \AA, about a factor of two higher than SDSS spectra. We
smoothed the spectra with a Gaussian kernel to match our working resolving power
${\sf R} \simeq 1000$. Fig. \ref{cflib} compares the atmospheric parameters in
the library's catalog, which were compiled from high-resolution studies in the
literature, with those we derived (FIT) for 333 stars in the effective
temperature range $5000 < T_{\rm eff} < 8000 $ K, discarding 54 stars which
could not be fit with a reduced $\chi^2< 20$. Our effective temperatures are in
good agreement with the catalog values. By fitting a Gaussian to the residuals
(see Fig. \ref{cflib}), we derive a mean offset of 1\%, and a 1$\sigma$ scatter
of 3\%. Our gravities are consistent with the catalog's, with a scatter of 0.30
dex, and an offset of $+0.02$ dex. Our iron abundances are also on the
same scale as those in the {\tt Cflib} catalog, 
with a scatter of only 0.16 dex. 

\subsection{{\sc elodie}}

\begin{figure}[t!]
\begin{center}
\includegraphics[angle=0,width=8cm]{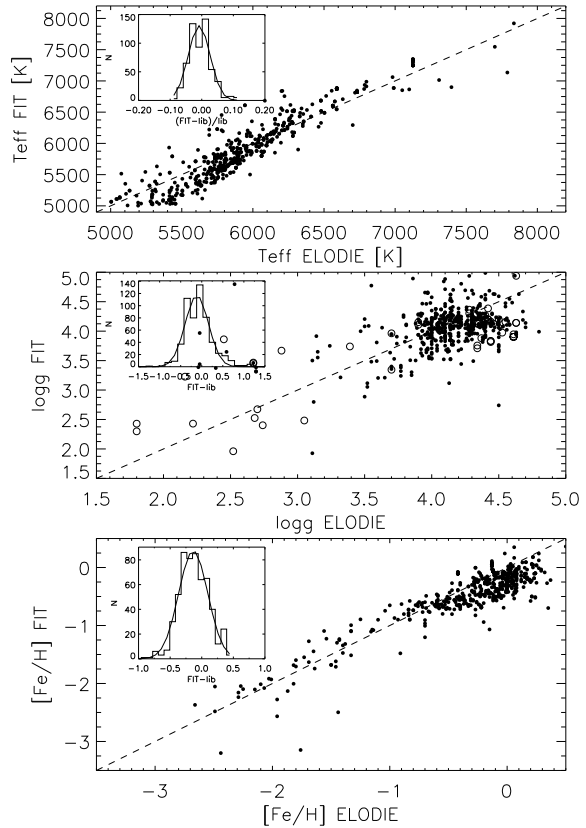}
\caption{Comparison between our derived metal abundances (FIT) and
those from the catalog of the {\sc elodie} library. 
Otherwise similar to Fig. \ref{cflib}.
 \label{elodie}}
\end{center}
\end{figure}

This library contains spectra obtained with the {\sc elodie} spectrograph at the
Observatoire de Haute-Provence 1.93m telescope, covering the wavelength range
400 to 680 nm. The version of the library employed here corresponds to 1969
spectra of some 1390 stars with a resolving power of ${\sf R} = 10,000$, publicly
available as part of the Elodie.3 release\footnote{http:
//www.obs.u-bordeaux1.fr/m2a/soubiran/elodie\_library.html} (Moultaka,
Ilovaisky, Prugniel, \& Soubiran 2004). We smooth the data with a Gaussian
kernel to our working resolution. Most spectra have originally a high $S/N$
ratio, which is subsequently enhanced by the smoothing process that we apply. 
The catalog values of
the stellar atmospheric parameters are those included in the FITS headers of the
spectra. They come, in most cases, from the literature. The FITS headers also
include a quality flag for the atmospheric parameters, and after noticing some
outliers with uncertain $T_{\rm eff}$s, we decided to limit the comparison to
stars with a quality flag in this parameter of at least 1, on the Elodie 0 to 4
scale. The atmospheric parameters for 567 spectra of stars with $5000 < T_{\rm
eff} < 8000$ K are compared with our values in Fig. \ref{elodie}. On average,
our effective temperatures are lower than Elodie's by 1\% ($\sigma=$ 4\%), our
gravities are lower by 0.12 dex ($\sigma =$ 0.28 dex), and our derived
abundances are lower by 0.14 dex ($\sigma=$ 0.23) dex. The systematic
differences in $T_{\rm eff}$ and [Fe/H] are not uniformly distributed, but
concentrate on the coolest and more metal-rich stars of the sample. By 
including cooler stars we are able to sample lower gravities. 
The open symbols in the
middle panel of Fig. \ref{elodie} correspond to 52 analyzed spectra with 
derived effective temperatures 4800 $< T_{\rm eff} <$ 5000 K, 
and suggest a scatter and systematic
offset very similar to those found for warmer stars with higher gravities
(filled circles).

After accounting for the systematic differences in $\log g$ and [Fe/H], our
derived $M_V$ magnitudes are systematically larger than those 
listed in the {\sc elodie} catalog. The median difference is 0.52 mag.

\subsection{\rm S$^4$N}
\label{s4nsection}

\begin{figure}[t!]
\begin{center}
\includegraphics[angle=0,width=8cm]{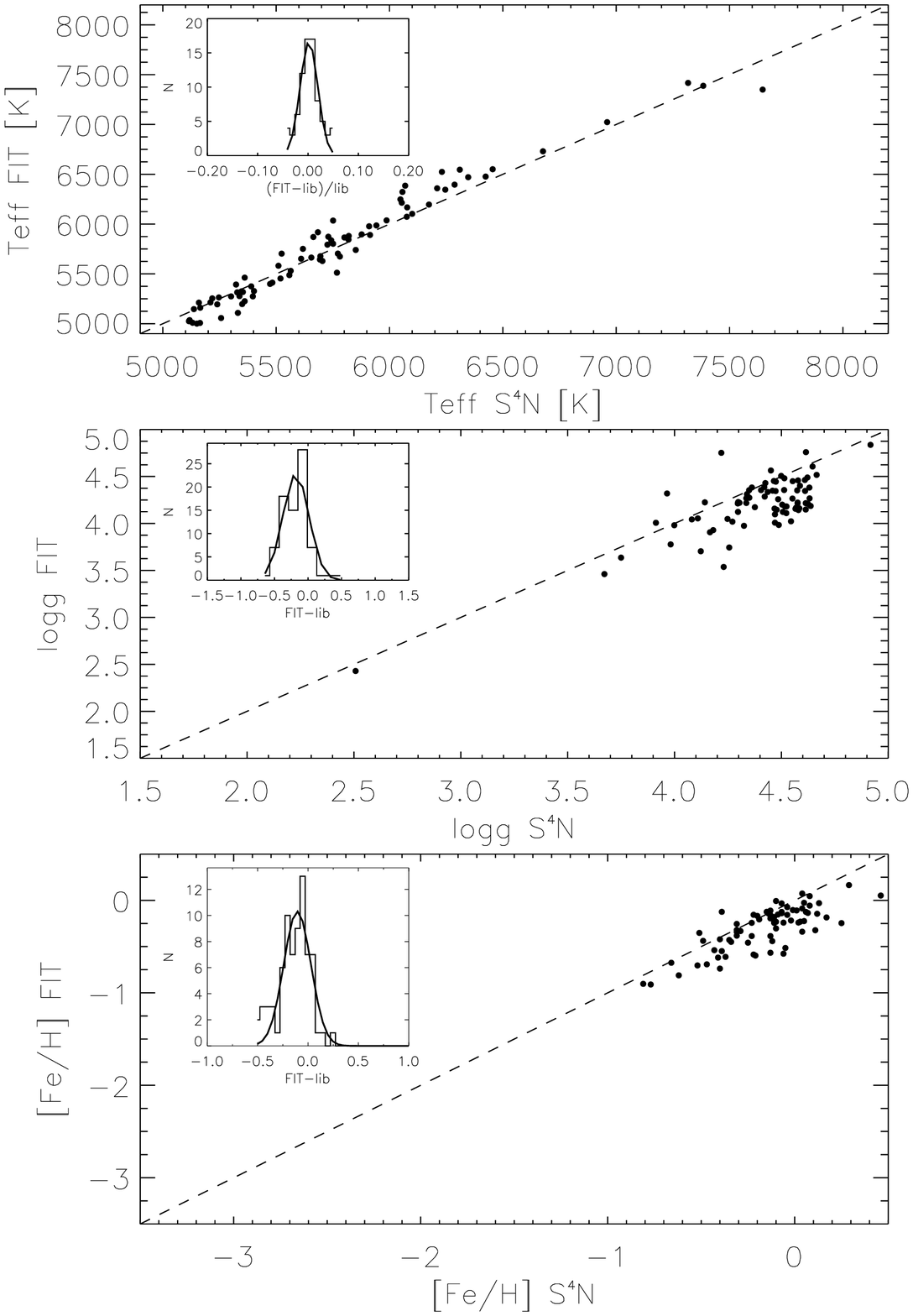}
\caption{Comparison between our derived metal abundances (FIT) and
those from the catalog of the S$^4$N library. 
Otherwise similar to Fig. \ref{elodie}. \label{s4n}}
\end{center}
\end{figure}

Our third comparison involves the S$^4$N survey, which includes spectra 
for all stars more luminous than $M_V=6.5$ mag within 14.5 pc from the Sun. 
These data were obtained with the 2.7m telescope at McDonald Observatory 
and the ESO
1.52m telescope at La Silla\footnote{{\tt http://hebe.as.utexas.edu/s4n/} ~~or
\\ {\tt http://www.astro.uu.se/$\sim$s4n/}}. This library consists of 118
spectra (for the same number of stars) with wide spectral coverage and a
resolving power of ${\sf R} \simeq 50,000$. 
Although much more limited in coverage of the
atmospheric parameters space than the previous two libraries, this data set is
useful because of the very high quality of both the spectra (the $S/N$ ratio
is usually larger than 300 at the original resolution) 
and the catalog of atmospheric
parameters. In this sample, there are 77 stars in the range $5000 < T_{\rm eff}
< 8000$ K. Fig. \ref{s4n} shows excellent agreement between the catalog and our
estimated atmospheric parameters for these nearby thin-disk stars. 
The $T_{\rm eff}$ scale
of S$^4$N is based on 
the Infrared Flux Method (IRFM; Blackwell \& Shallis 1977). 
The average offset is insignificant, and the 1$\sigma$ scatter is 2 \%,
although there is a slight trend with temperature. The gravities, based on {\it
Hipparcos} parallaxes for the S$^4$N catalog, are again offset, with our values
lower by $0.18$ dex, with a scatter of 0.20 dex. The familiar star Capella, with
$\log g \sim 2.5$, is the only giant in the sample. The metallicity comparison
is again excellent, with our values lower by $0.11$ dex, and a scatter of 0.14
dex.

\begin{figure*}[t!]
\begin{tabular}{cc}
\includegraphics[angle=90,width=8cm]{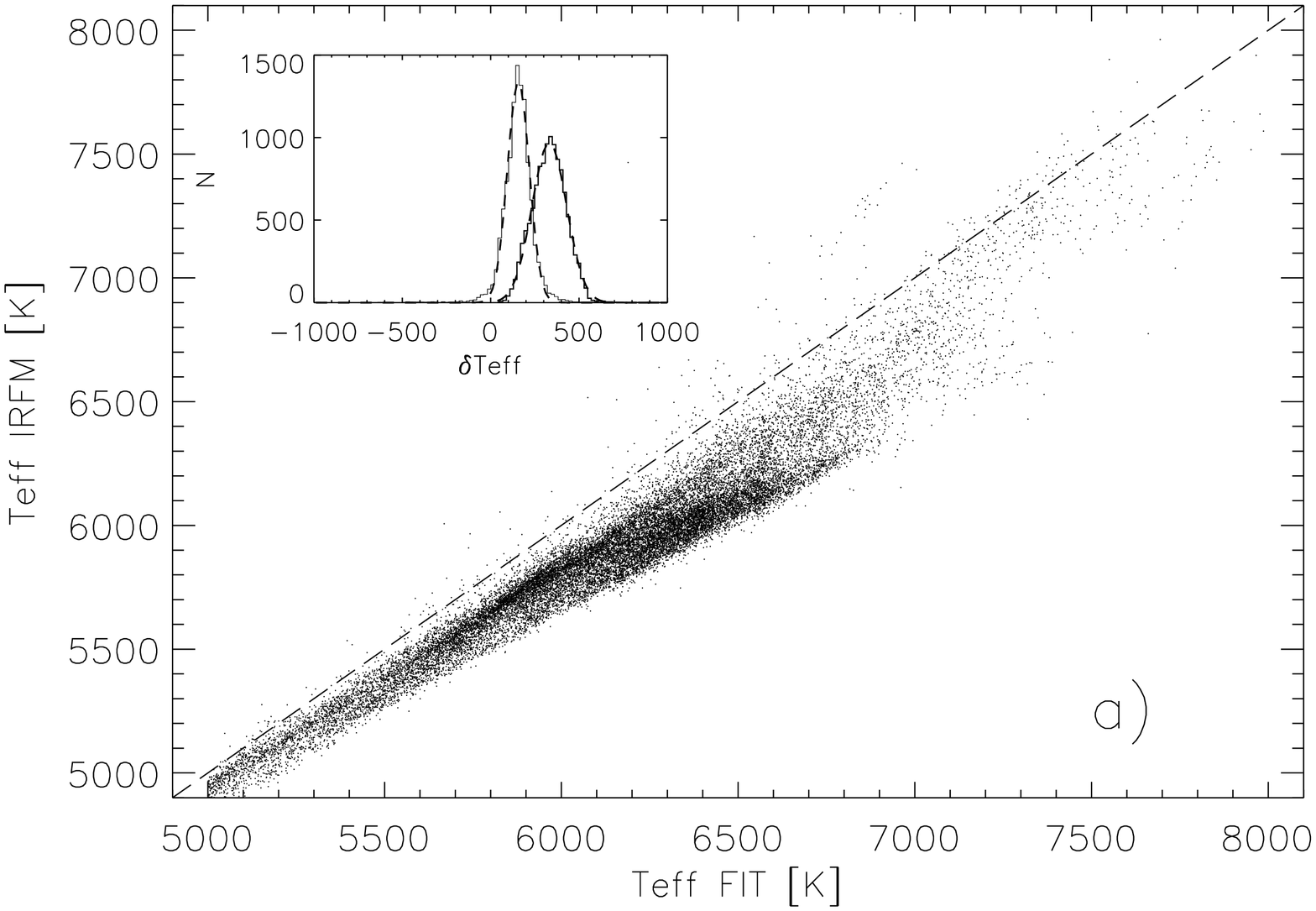} &
\includegraphics[angle=90,width=8cm]{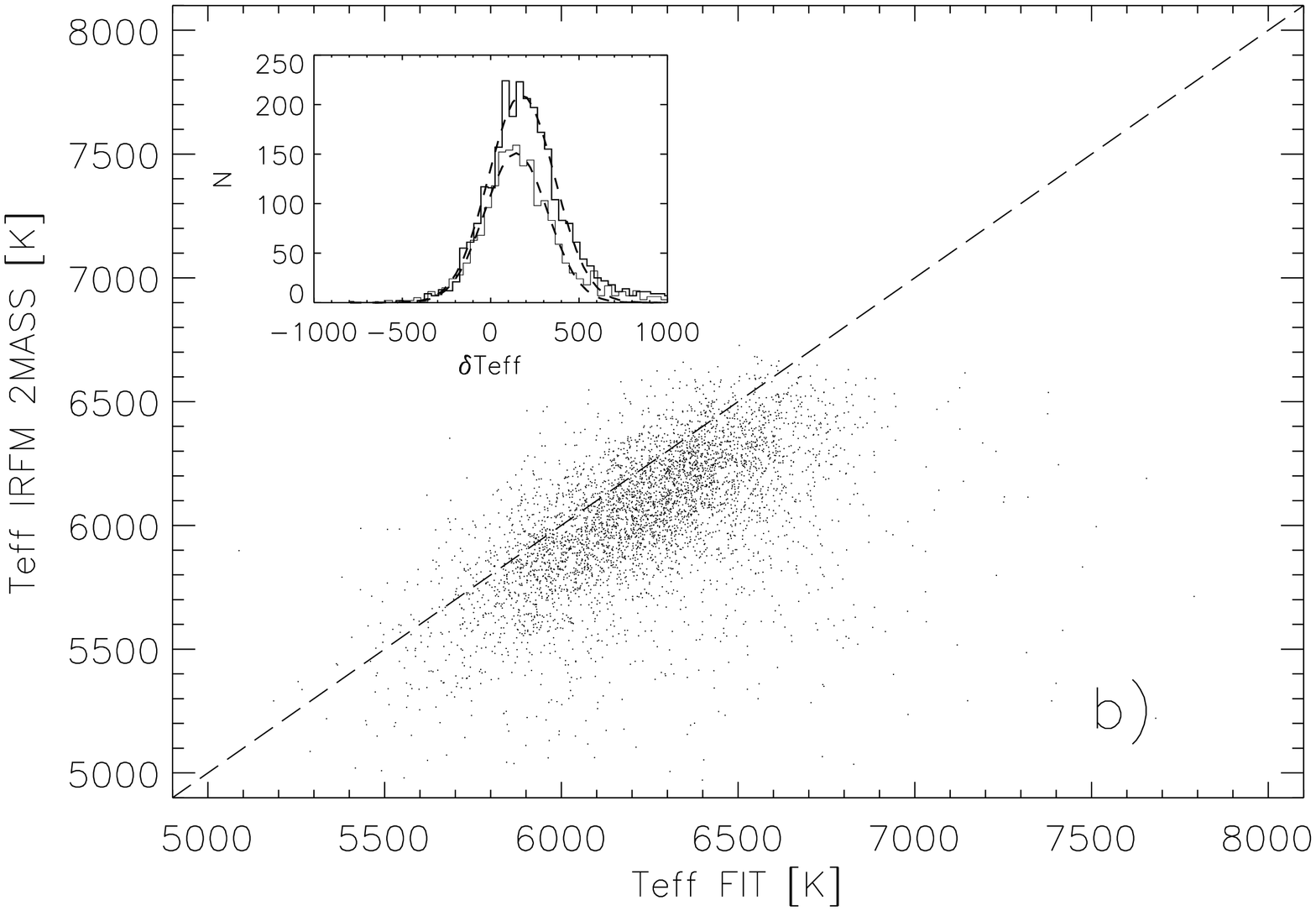} \\
\includegraphics[angle=90,width=8cm]{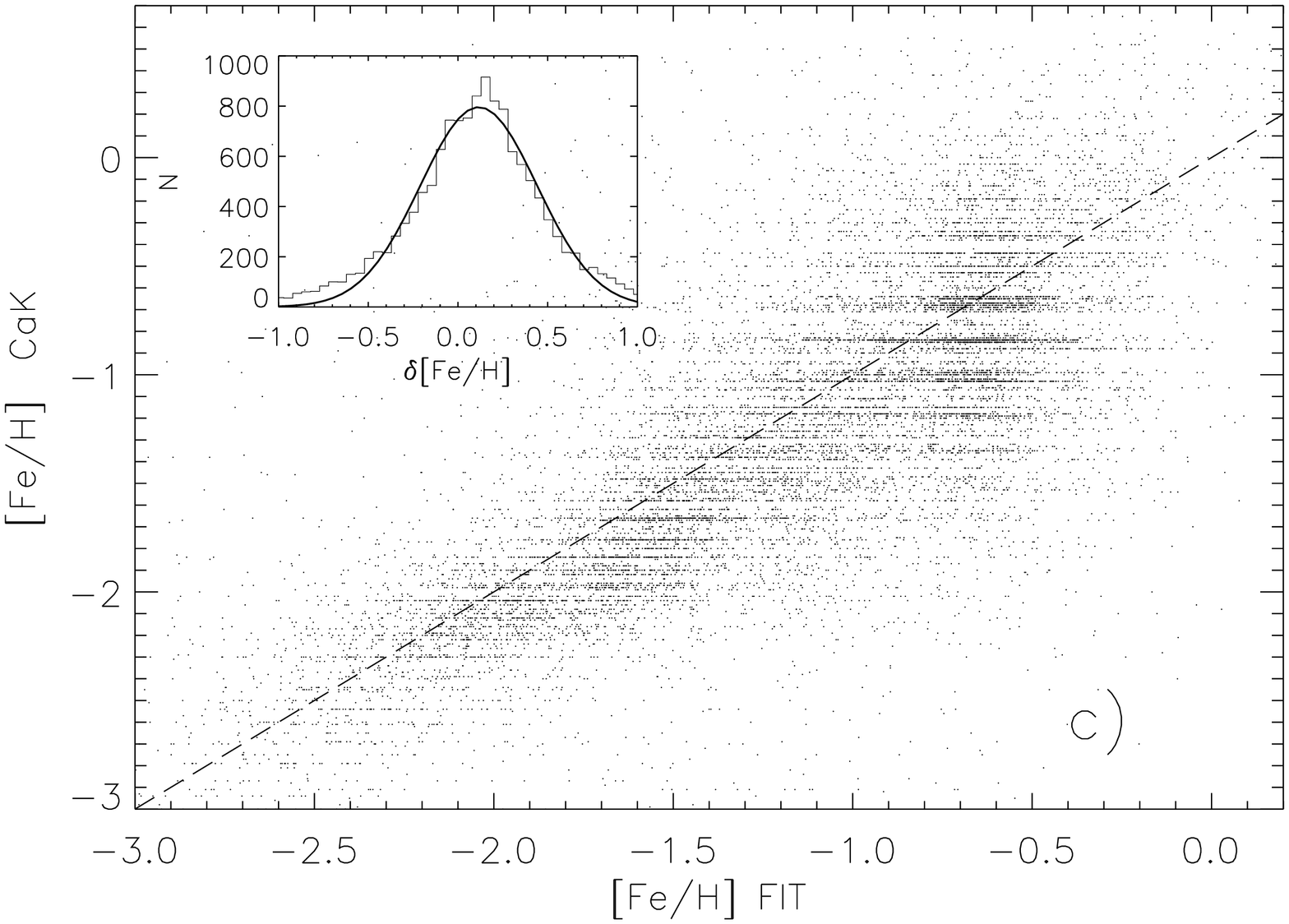} & 
\includegraphics[angle=90,width=8cm]{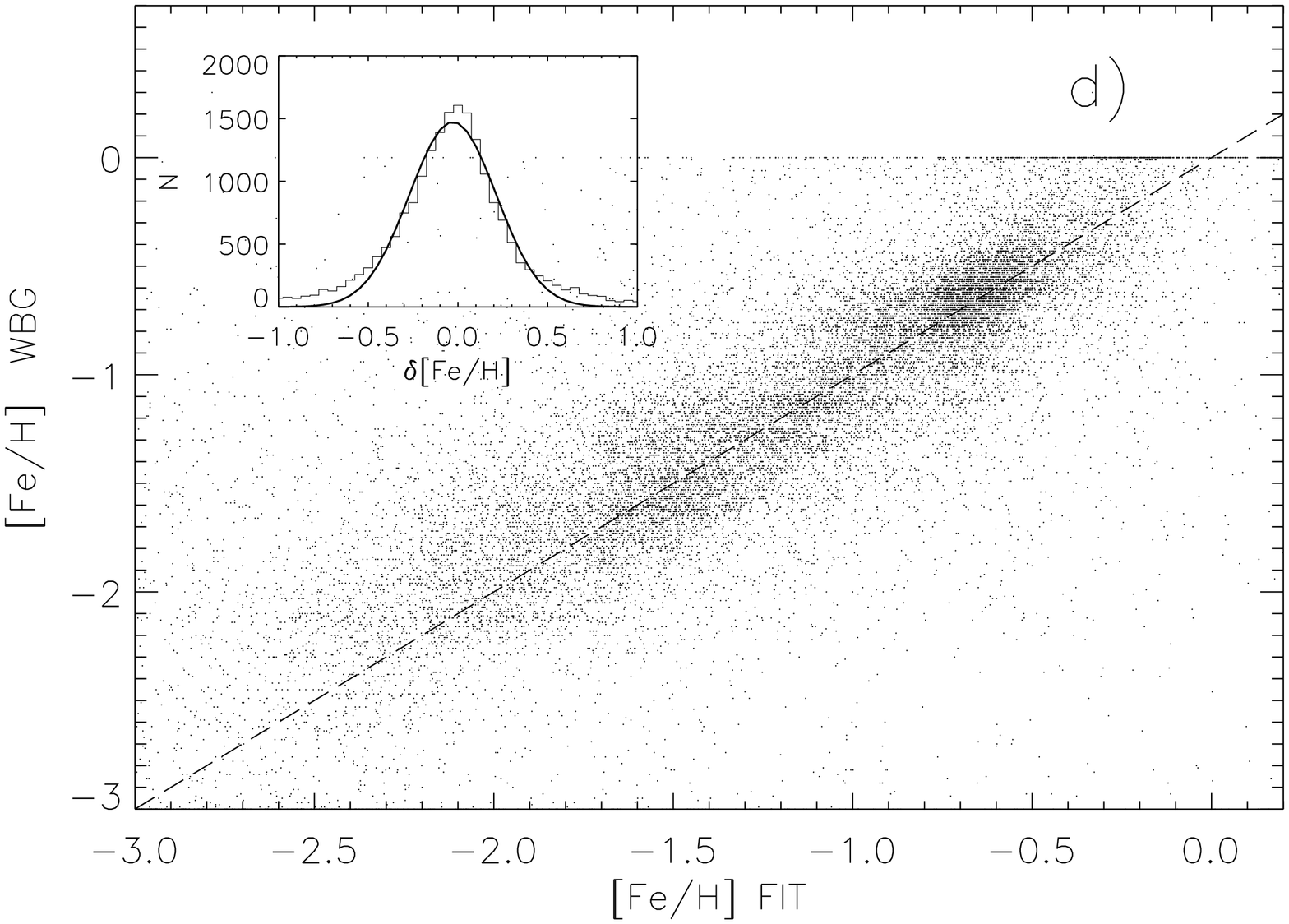} \\
\end{tabular}
\caption{
{\it a)}: Comparison between our derived effective temperatures
and those from calibrations of optical color indices based on the infrared
flux method; {\it b)} Similar to {\it a)}, but for color indices that 
employ 2MASS observations; 
{\it c) and d)}: Comparison between our derived metal abundances (FIT) and
those from the `CaK' and `WBG' methods (see Beers et al. 1999 and Wilhelm 
et al. 1999a, respectively). 
The dashed line has a slope of
unity. The inset graphs show the differences (FIT - other; histograms) 
and a least-squares Gaussian fit, with the sample divided
depending on metallicity at [Fe/H]$=-1$ for panels 
{\it a)} and {\it b)}. \label{cak}}
\end{figure*}

The ultra-precise parallaxes from Hipparcos available for this sample
(1 \% on average) offer a good opportunity to evaluate the quality of the
absolute magnitudes and distances obtained from the comparison with stellar
evolution calculations (see \S \ref{ana}). After correcting our surface
gravities for the bias mentioned above, we find 
that our $M_V$ values are systematically
larger than those derived from the Hipparcos parallaxes by 0.5 mag ($\sigma =
0.4$ mag), confirming the offset inferred from the analysis of {\sc elodie}
spectra. On average, the derived distances $\log_{10} d = (5 + V - M_V)/5$ are
21 \% smaller than the inverse of the parallaxes ($\sigma= 19 \%$), with no
obvious dependence on the atmospheric parameters.

\subsection{The effect of $S/N$}

Our previous comparisons with existing libraries of {\it nearby} stars helped to
established zero points of our scale of atmospheric parameters, as well as to
estimate the uncertainties involved in the analysis of real observations.
Nonetheless, the spectra included in these libraries have a quality 
that is far higher than typical SDSS observations. 
To estimate realistic values for the
uncertainties in the derived parameters, we introduce noise in the
observations and re-analyze them.

We chose to experiment with spectra from {\tt Cflib}, which are those originally
more similar to the SDSS observations. By introducing Poisson noise in the
spectra, we degraded them to a $S/N$ per pixel\footnote{Our resolution element
includes 3 pixels.} of 160, 80, 40, 20, and 10. We also introduced noise in the
$(g-r)$ index, according to the empirical relationship described in \S
\ref{ana}. The results of this experiment are included in Table \ref{t1}, and
suggest that our analysis procedure, and in particular the metallicity
determination, is quite robust to reductions in signal-to-noise ratio.

\begin{deluxetable*}{ccrccccc}[b!]
\tablecaption{Offsets between our derived atmospheric parameters
and the catalog values}
\tablenum{1}
\tablehead{
\colhead{} & \colhead{} & \multicolumn{2}{c}{$T_{\rm eff}$} & \multicolumn{2}{c}{$\log g$} & \multicolumn{2}{c}{[Fe/H]} \\
\colhead{Library} & \colhead{$S/N$\tablenotemark{1}} & \colhead{$<\Delta>$} & \colhead{$\sigma$} & \colhead{$<\Delta>$} & \colhead{$\sigma$} & \colhead{$<\Delta>$} & \colhead{$\sigma$} \\ 
\colhead{} & \colhead{} & \colhead{(\%)} & \colhead{(\%)} & \colhead{(dex)} & \colhead{(dex)} & \colhead{(dex)} & \colhead{(dex)} } 
\startdata
{\sc elodie} & Full & $-0.8$ & 3.5 & $-0.12$ & 0.28 & $-0.14$ & 0.23 \\
S$^4$N       & Full & 0.2    & 1.8 & $-0.18$ & 0.20 & $-0.11$ & 0.14 \\
{\tt Cflib}  & Full & 1.3    & 3.4 & $+0.02$ & 0.27 & $-0.02$ & 0.16 \\
{\tt Cflib}  & 160  & 1.1    & 3.8 & $-0.05$ & 0.35 & $-0.02$ & 0.15 \\
{\tt Cflib}  & 80   & 1.3    & 3.9 & $-0.04$ & 0.33 & $-0.02$ & 0.16 \\
{\tt Cflib}  & 40   & 0.8    & 3.4 & $-0.01$ & 0.43 & $-0.06$ & 0.20 \\
{\tt Cflib}  & 20   & 0.8    & 4.8 & $+0.00$ & 0.51 & $-0.09$ & 0.31 \\
{\tt Cflib}  & 10   & 1.6    & 8.3 & $+0.06$ & 0.75 & $-0.13$ & 0.46 \\
\enddata
\label{t1}
\tablenotetext{1}{The $S/N$ values here considered correspond to 1 pixel, or
1/3 of a resolution element.}
\end{deluxetable*}

\subsection{Empirical corrections}

The offsets in surface gravity and metallicity found in the comparison with
S$^4$N and {\sc elodie} are very similar, and therefore we conclude that
corrections of about $+0.15$ dex and $+0.12$ dex should be applied to our
gravities and iron abundances, respectively, to match the scale of these
libraries. Nevertheless, these libraries are mainly comprised of stars with high
metallicities. Closer inspection of the {\sc elodie} analysis shows that the
metallicity offset disappears when the comparison is limited to stars with
[Fe/H]$<-1$; this is also supported by the comparison with {\tt Cflib}.
Thus, we choose not to correct for this possible offset for the
DR3 sample. The offset in surface gravity persists at all metallicities, hence
we apply this correction to the values derived in the analysis of 
DR3 data. 

Another important conclusion that emerged from comparison with the {\sc
elodie} and S$^4$N samples is that our derived $M_V$s appear to be 
offset by $\simeq 0.5$ mag. This effect may decrease for lower gravities, but
because it is clear for dwarfs and subgiants, which make up most of our SDSS
sample, we correct for it in our subsequent analysis of DR3 stars. We 
emphasize that this is equivalent to a correction in the derived distances of 
about 20 \%. As shown below, systematic errors of this order in the stellar 
distances are inconsequential for our main conclusions. 

As can be readily noticed in Figs. \ref{cflib}, 
\ref{elodie},  and \ref{s4n}, the number of stars with temperatures in the range
$7000< T_{\rm eff} < 8000$ K is very limited. Nevertheless, analysis of the
gravities for the {\sc elodie} sample for the few stars in this regime suggests
that our surface gravities are overestimated by $+0.4$ dex, instead of
underestimated by $\sim 0.12$ dex as previously found for the overwhelming
majority of stars with cooler temperatures. To avoid poorly understood 
systematic errors, we restrict the following discussion to the temperature range
$5000 < T_{\rm eff}< 7000$ K, approximately equivalent to spectral types
F and G.

\subsection{Comparisons with other methods on SDSS data}
\label{other}

Using Eq. \ref{zn} we can easily estimate $B-V$ colors for the DR3 stars
analyzed and, by combining this information with the spectroscopically
determined surface gravities and metallicities, apply the IRFM calibrations of
Alonso et al. (1996, 1999) to estimate effective temperatures. This exercise
reveals an intriguing trend, which is illustrated in Fig. \ref{cak}a. Our
derived temperatures (labeled as {\sc fit} in the Figure) are systematically
higher by an amount that increases with $T_{\rm eff}$. The inset shows the
distribution of the differences for two subsets of the sample. For stars with a
metallicity [Fe/H] $> -1$, the mean offset is 162 K ($\sigma = 62$ K), but for
[Fe/H] $ < -1$, the average difference increases to 337 K ($\sigma=109$ K).

Offsets on the order of 100--150 K between the IRFM calibrations and 
 effective temperatures derived from spectroscopy 
 have been repeatedly reported in the
literature for disk stars (e.g., Santos, Israelian, \& Mayor 2004; Yong et al.
2004; Luck \& Heiter 2005; Takeda et al. 2005). There also exists an indication
of a similar effect for F-type stars in the upper panel of Fig. \ref{s4n}.
However, the enhanced discrepancies at lower metallicities raise some concern
about the use of the $(g-r)$ transformation in Eq. \ref{zn} for such stars. We
note that our analysis of SDSS data is independent from this transformation, but
it plays a role in the comparison with the reference libraries. A transformation
that differs somewhat from Eq. \ref{zn} has been recently proposed by Bilir,
Karaali,\& Tun\c{c}el (2005)

\begin{equation}
(g-r) = 1.12431 (B-V) -0.25187,
\label{bilir}
\end{equation}

\noindent but our tests revealed that the use of this relation instead of 
Eq. \ref{zn} would only have a very small effect on the comparison 
with the libraries
in the previous sections. Interestingly, the metallicity-dependent discrepancies
shown in Fig. \ref{cak}a would be significantly enhanced by the use of Eq.
\ref{bilir}; the same is true if we use instead the transformations derived
for the $u'g'r'i'z'$ system by Smith et al. (2002).

By comparison of our sample with the 2MASS All Sky Catalog (Cutri et al.
2003\footnote{\tt http://www.ipac.caltech.edu/2mass/index.html}), we were able
to identify 10,210 sources with available 2MASS photometry ($JHK_s$). The IRFM
calibrations recently derived by Ram\'{\i}rez \& Mel\'endez (2005) consider the
indices $(V-J)$, $(V-H)$ and $(V-K_s)$, based on 2MASS $JHK_s$ photometry.
Unfortunately, applying these calibrations still requires an intermediate step,
in order to derive $V$ from $g$. We can resort to another of the relations
derived by Zhao \& Newberg:

\begin{equation}
V= g -0.561(g-r) -0.004,
\label{zn2}
\end{equation}
\noindent or to a combination of Eqs. (1) and (3) of Bilir et al.:
\begin{equation}
V = g - 0.56353(g-r) -0.03381.
\label{bilir2}
\end{equation}

\noindent For about 4,000 stars, mostly dwarfs in the range 
$5500 < T_{\rm eff}< 6500$ K, the IRFM calibrations involving the three 2MASS
passbands (ignoring reddening) yield temperatures that agree within 200 K. For
those stars, we find a mean systematic offset between our effective temperatures
and the IRFM calibrations of 235 K, when Eq. \ref{zn2} is used, or 173 K when
adopting Eq. \ref{bilir2}. The distribution of residuals is very similar in both
cases, and well approximated by a Gaussian with $\sigma= 180$ K. There is only a
minor excursion from the mean offset depending on metallicity (of about 40 K for
stars above and below [Fe/H]$=-1$, respectively), as illustrated in Fig.
\ref{cak}b for the case when Eq. \ref{bilir2} is used. Furthermore, the offset
is largely independent of $T_{\rm eff}$, and thus approximately equivalent to a
correction of $\sim -0.1$ to the $V$ magnitudes derived from the equations
above. Because the Ram\'{\i}rez \& Mel\'endez calibrations are in good
correspondence with the temperature scale of the Alonso et al. calibrations, and
all our stars are far enough to make distance-dependent extinction corrections
implausible, we conclude that the most likely explanation for the systematics in
Fig. \ref{cak}a has to do with the photometric transformations between Johnson
and Sloan color indices. Clearly, further work on the derivation of suitable
color transformations, especially as a function of metallicity, is required.

\begin{figure*}[t!]
\includegraphics[angle=90,width=16.5cm]{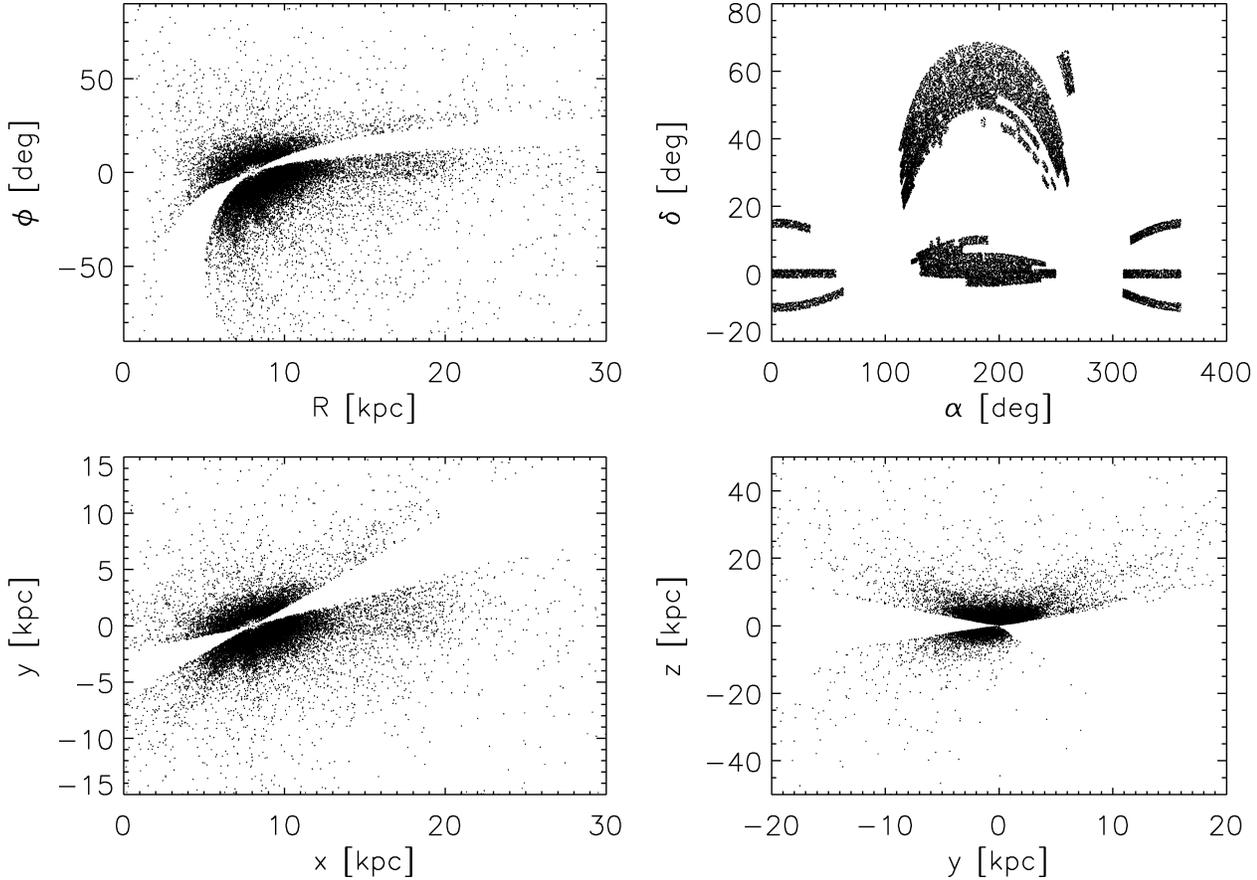}
\caption{Projected positions of the DR3 stars selected for analysis
on the galactic plane (top left-hand panel), on the sky (top right), 
and on galactic cartesian coordinates (bottom panels). \label{pos}}
\end{figure*}

\begin{figure}[h!]
\includegraphics[angle=90,width=8cm]{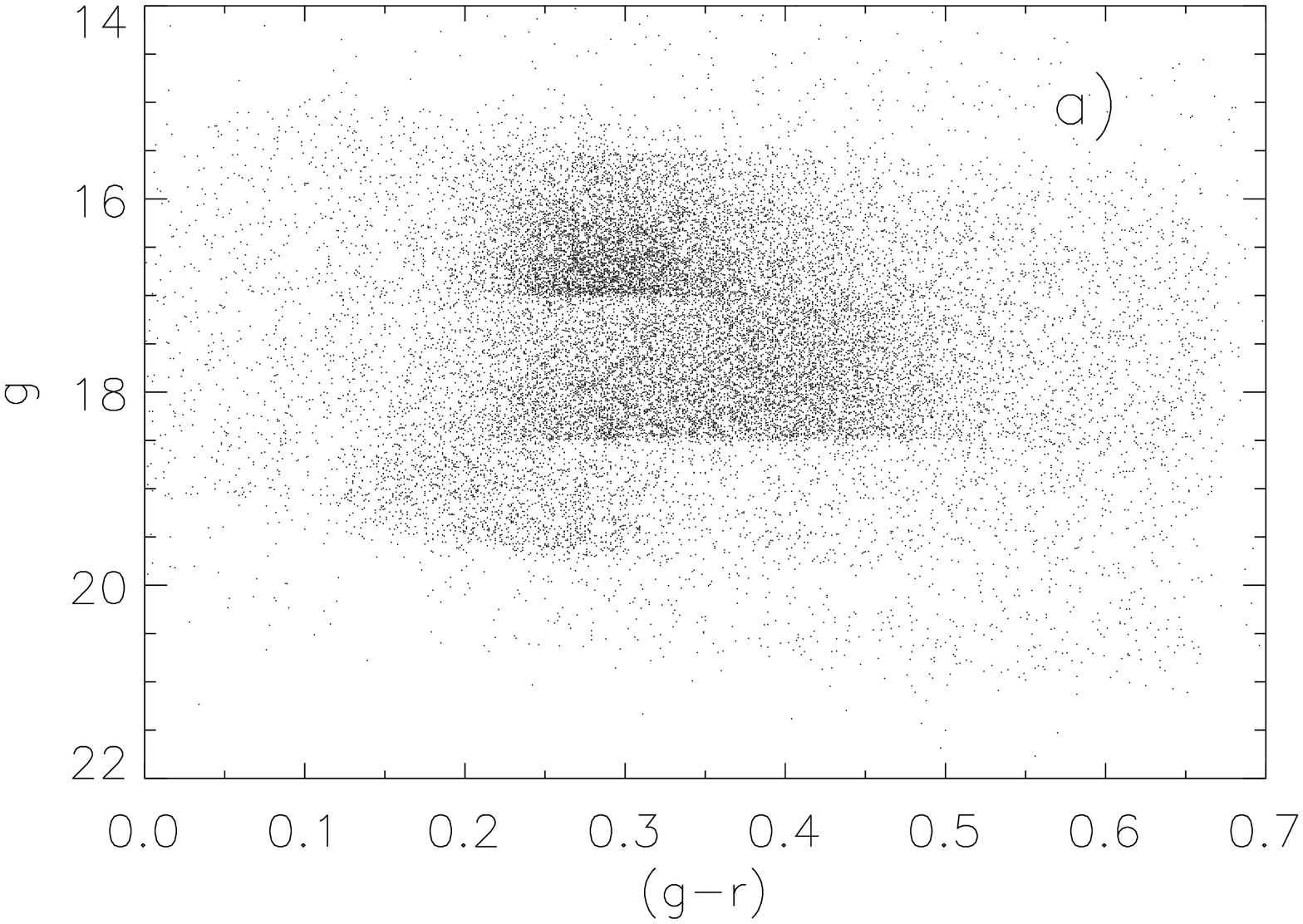}
\includegraphics[angle=90,width=8cm]{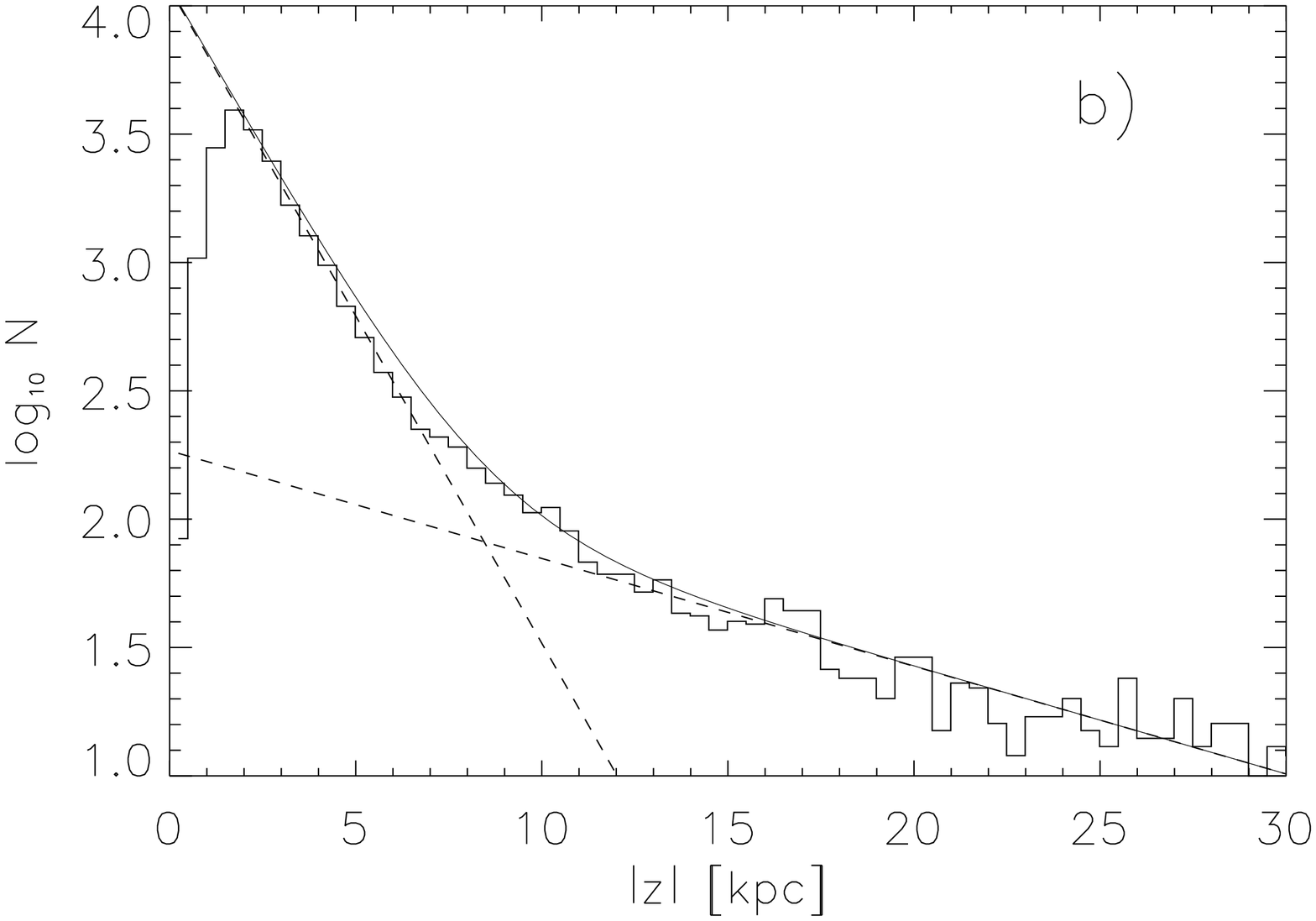}
\caption{{\it a)}: Distribution of the stars selected for analysis on the
$g-(g-r)$ plane. The large numbers of F-type subdwarfs
selected as flux calibrators are obvious in the range 
$0.2<(g-r)<0.5$,$15.5<g<18.5$. {\it b)}: 
Number of stars in the sample as a function of distance from the 
galactic plane. The stars counts are reasonably well-fit with a combination
of two exponential laws (dashed lines). \label{grg}}
\end{figure}

We have also compared our iron abundances for stellar spectra in DR3
with the values derived from two different techniques, one based on the
equivalent width of the Ca II K and an estimated $B-V$ color (`CaK'; see Beers
et al. 1999), and a second based on $ugriz$ photometry and the equivalent width
of the Ca II K line, supplemented with spectral synthesis of Balmer lines and
molecular bands in the blue region of the spectrum (`WBG'; a refined version of
the method described by Wilhelm, Beers \& Gray 1999a). The Ca II K line
approaches saturation at high metallicity and cooler temperatures, but it should
be a reliable indicator at low metallicity. Fig. \ref{cak}c confronts the two
scales for 18,727 DR3 stars with $(B-V) < 1$, or spectral types earlier than K0.
On average, our iron abundances are higher than those obtained from the CaK
method by $0.12$ dex, with a $1\sigma$ scatter of 0.33 dex, which is reduced for
stars with [Fe/H]$<-1$. Over much of the range of [Fe/H], the comparison with
the WBG values for 20,990 stars shows less scatter than with the CaK method,
only 0.24 dex, and no detectable offset, as illustrated in Fig.
\ref{cak}d.  However,  the scatter in the estimated [Fe/H]
is lower when compared with the CaK approach at the lowest metallicities.

Our derived distances can be compared with those obtained purely from
photometry, using the main-sequence relationship (Bilir et al. 2005)
\begin{equation}
M_g = 5.791 (g-r) + 1.242 (r-i) + 1.412.
\label{bilir3}
\end{equation}
On average, the distances based on the spectroscopic parameters 
are $0.33$ dex (or 76 \%) smaller than the photometric version, with
a 1$\sigma$ scatter of about 20 \%. The comparison in \S \ref{s4nsection}
showed that our inferred distances  were biased by only $\sim 20$ \%,
and therefore we believe most of the discrepancy with the 
photometric distances is likely related to issues with the latter.

\begin{figure*}[t!]
\begin{tabular}{cc}
\includegraphics[angle=0,width=9cm]{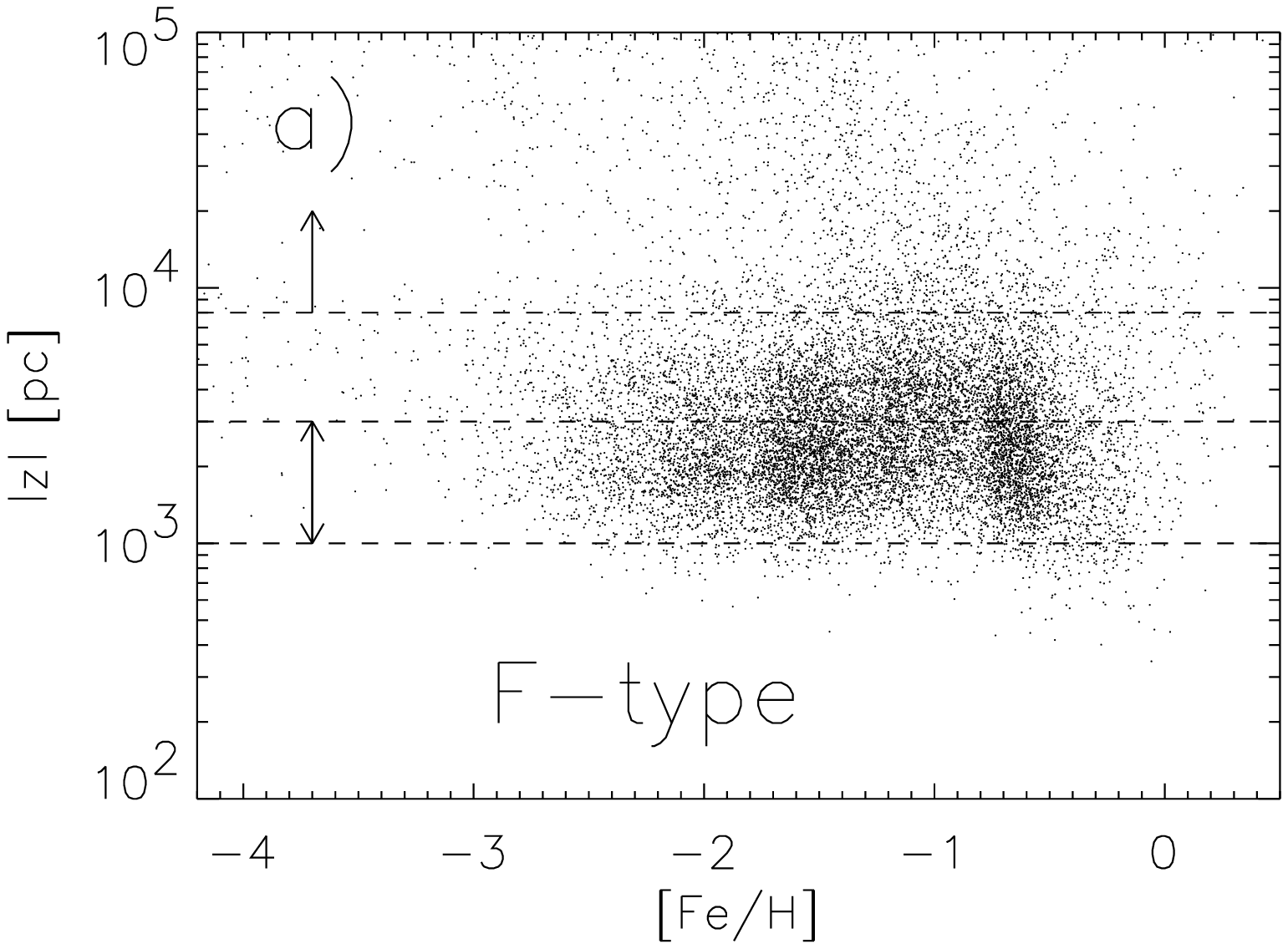} & 
\includegraphics[angle=0,width=9cm]{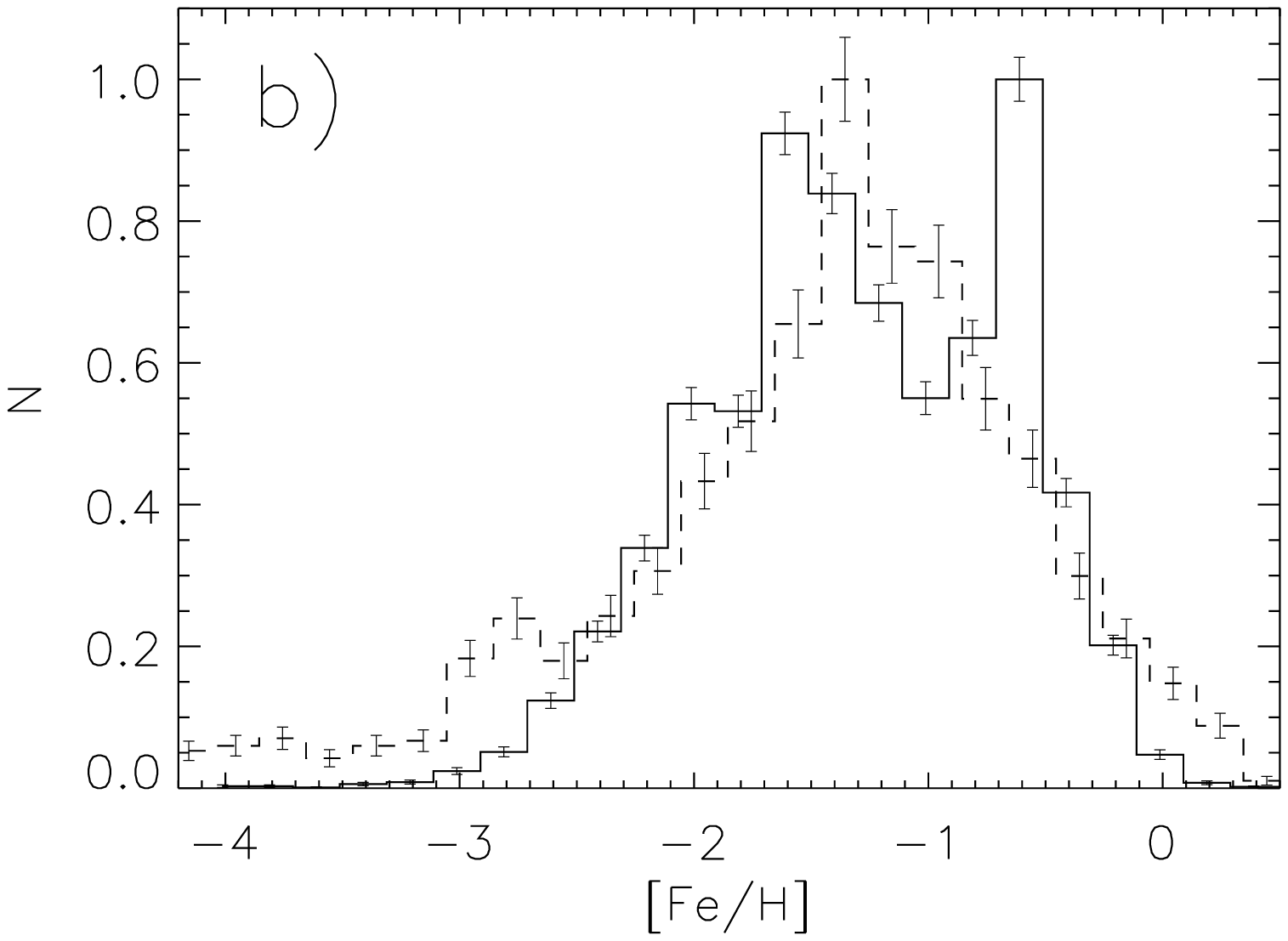} \\
\includegraphics[angle=0,width=9cm]{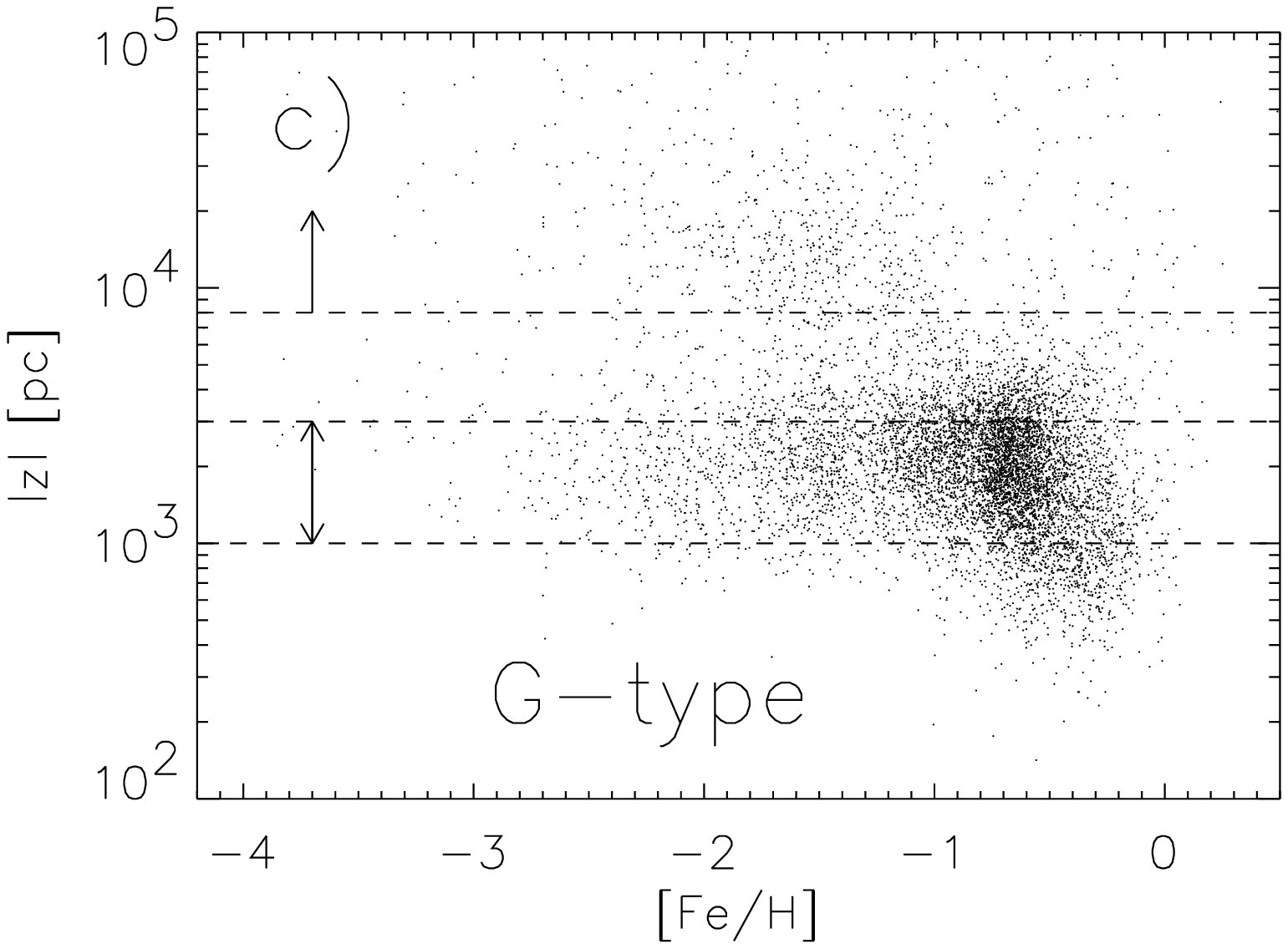} & 
\includegraphics[angle=0,width=9cm]{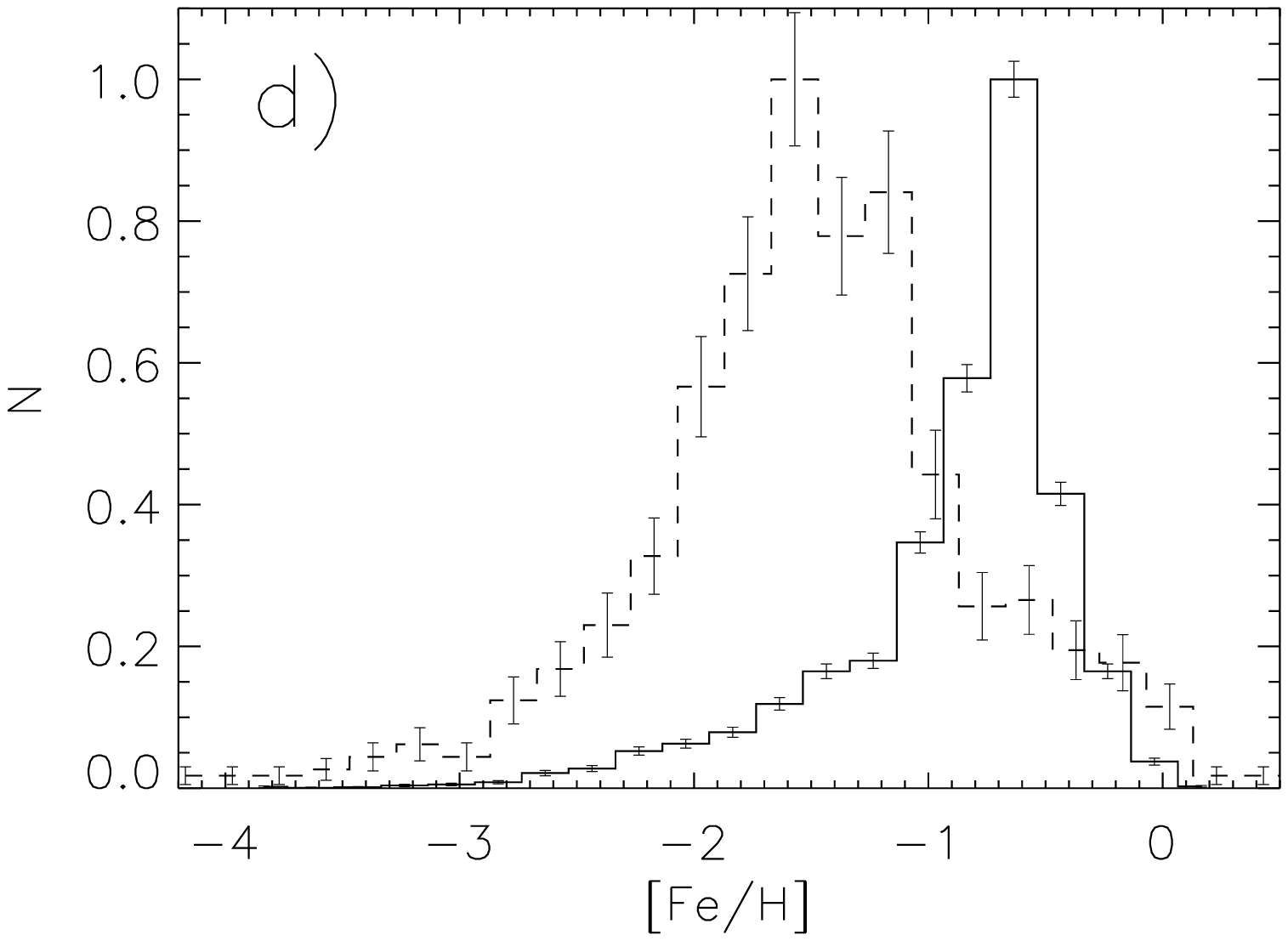} \\
\end{tabular}
\caption{{\it a)} 
Metallicity distribution of the F-type stars in the sample 
(6000$<T_{\rm eff}< 7000$ K)
as a function of
distance from the galactic plane; {\it b)} The stars in panel a) between 
1 and 3 kpc from the plane (solid) and those farther away than 8 kpc 
(dashed line) 
are binned in 0.2 dex intervals to calculate the histograms shown. 
There are 7,422 stars at $|z|<3$ kpc and 2,107 stars at $|z| > 8$ kpc;
{\it c) and d)} Similar to a) and b), but only including G-type stars 
(5000$<T_{\rm eff}< 6000$ K).
There are 5,061 stars at $|z|<3$ kpc and 731 stars at $|z| > 8$ kpc.
\label{mdf}}
\end{figure*}

\subsection{Duplicates}

DR3 includes some instances of multiple observations of the same stars
in different plugplates. Among the 44,175 DR3 spectra for which we successfully
identified Balmer lines, there are 1,022 duplicates, 11 objects with three
spectra, and 1 star with 4 observations. For these cases, the $S/N$ ratio of the
spectra were very similar, given the uniform exposure times used in SDSS,
and thus we retained the averaged values for the stellar parameters. 
The parameters derived from multiple spectra of the same object 
are remarkably consistent The
mean rms scatter (basically the average difference, 
for most of the multiple
observations are just duplicates) 
in $T_{\rm eff}$, $\log g$, and [Fe/H] are 16 K, 0.09 dex, and 0.07 dex, 
respectively.

\section{Results from DR3}
\label{dr3}

Fig. \ref{pos} shows the projected positions of the selected SDSS stars on the
sky (equatorial coordinates {$\alpha$,$\delta$}) on the galactic plane
(polar coordinates {$R$, $\phi$}), where

\begin{equation}
\label{rgaleq}
 R =   \displaystyle \sqrt{(R_{\odot}- d \cos b \cos l)^2 + (d \cos b \sin l)^2},
\end{equation}

\begin{eqnarray}
\label{phi}
\displaystyle \sin \phi = & \displaystyle - \frac{d \cos b \sin l}{R}, ~~~{\rm and}\nonumber\\
		 & 	\nonumber\\
\displaystyle \cos \phi = & \displaystyle  \frac{R_{\odot} - d \cos b \cos l}{R}, \nonumber\\
\end{eqnarray}

\noindent and in cartesian galactic coordinates ($x = R \cos \phi$, 
$y= R \sin \phi$, $z= d \sin b$). The solar position in the top left-hand panel is at $\phi=0$ 
and $R_{\odot}$, the galactocentric distance of the Sun, which is taken as 8
kpc. The gaps in coverage on the galactic plane result from those 
of the DR3 footprint on the sky.

The selection effects involved in choosing SDSS spectroscopic targets are
complex, and we have imposed additional cuts in effective temperature. Our final
sample includes 22,770 F- and G-type stars with $5000 < T_{\rm eff}<7000$ K and
$S/N > 30$; their distribution in the $g$ vs. $(g-r)$ plane is shown in
Fig. \ref{grg}a. Some patches in the density of stars are obvious, such as the
16 stars per plate targeted, in similar proportions, 
as flux calibrators ($g<17$, $0.<g-r<0.6$) or  
reddening {\it standards} ($g<18.5$, $0<g-r<0.6$). These standards
are also subjected to two more color cuts
%
$0.6  <  u-g  <1.2$ and $g-r >   0.75 (u-g)  -0.45$, 
which introduce a bias against high-metallicity
([Fe/H] $> -0.5$)
stars. In addition, standard-star candidates are prioritized based on
the proximity of their colors to those of the SDSS standard halo subgiant 
BD $+17$ 4708, which strengthens the bias.
The density of stars at the red end is
artificially enhanced in DR3, as brown-dwarf candidates were given a high
priority, but they have been excluded from Figure \ref{grg} 
for being outside the
domain that our analysis can handle. A similar bias applies to A-type
horizontal-branch candidates with $(g-r)<0.2$. The distribution of DR3
spectroscopic targets resembles the distribution of photometric point sources
for $16.5 < g < 18.5$ and $0.5<(g-r)<0.85$, but unfortunately the distance
range covered by the brightest dwarfs in that range is limited to $1 < d < 3$
kpc, which is insufficient to properly map the falloff of the stellar density of
the thick disk.

\subsection{Metallicity distribution}
\label{metal}

Fig. \ref{grg}b shows the number of stars as a function of distance 
from the
galactic plane, $z= d \sin b$, assuming the Sun is at $z=0$. The combination of
two exponential laws provides a reasonable approximation for the star counts.
Even though we cannot properly derive the stellar density of the Galaxy from the
star counts, we can certainly associate, based on previous work, the majority of
the stars within $1<|z|<3$ kpc with the thick-disk 
population and virtually all of those at
$|z| > 8$ kpc with the halo population. 
More specifically, assuming scale heights
$Z^h_{\rm thin} \simeq 0.25$ kpc and $Z^h_{\rm thick} \simeq 0.8$ kpc, and a
normalization factor $\rho_{\rm thick}/\rho_{\rm thin} \sim 0.1$, more than
83 \% of the stars at $1<|z|<3$ kpc are members of the thick disk; this
number would be enhanced if metal abundances are considered. Adopting a
spherical halo whose density falls from the galactic center as a
power-law with an exponent of $-2.44$ (Robin et al. 2003), and a 
ratio $\rho_{\rm thick}/\rho_{\rm halo} \sim 40$ at the solar position,
we find that more than 99 \% of the stars at $|z|>8$ kpc are members of 
the halo.

The sample considered here contains 12,483 stars at $1<|z|<3$ kpc and 2,838 at
$|z| > 8$ kpc. Figure \ref{mdf}a shows the metallicities and distances from the
plane for individual F-type stars ($6000 < T_{\rm eff}< 7000$ K). Histograms of
the metallicities of the stars in these two $|z|$ ranges are shown in Fig.
\ref{mdf}b (solid: $1<|z|< 3$ kpc; dashed: $|z|> 8$ kpc). Figs. \ref{mdf}c and d
are the equivalent diagrams for G-type stars ($5000 < T_{\rm eff}< 6000$ K). No
significant differences are found between the metallicity distribution function
 of stars in the North or South galactic hemispheres. 

Some of the most metal-rich stars in the sample at $|z|<1$ kpc are most likely
members of the thin-disk population. As was previously found for EDR spectra,
dwarfs and subgiants with SDSS spectra in DR3 trace the thick disk (with some
contamination from the thin disk), while the brighter and more evolved stars
belong mostly to the halo (Allende Prieto et al. 2004b). The halo metallicity
distribution, as derived from the stars at $|z|> 8$ kpc, peaks approximately at
[Fe/H] $\sim -1.4$, and contains stars with a wide range of metallicities. The
peak of the metallicity distribution function of the thick disk (metal-rich
peaks at $1<|z|<3$ kpc) is located at about [Fe/H] $\simeq -0.7$, in good
agreement with previous determinations (e.g., Gilmore, Wyse, \& Jones 1995). The
distribution of F-type stars close to the plane shows a second-peak roughly at
the same metallicity of the distant halo stars. We expect that some of these
stars belong to the metal-weak thick disk, (see Beers et al. 2002 and references
therein), but most do not share the kinematics of the more metal-rich thick-disk
stars (see \S \ref{kin}).

\begin{figure}[t!]
\includegraphics[angle=0,width=8cm]{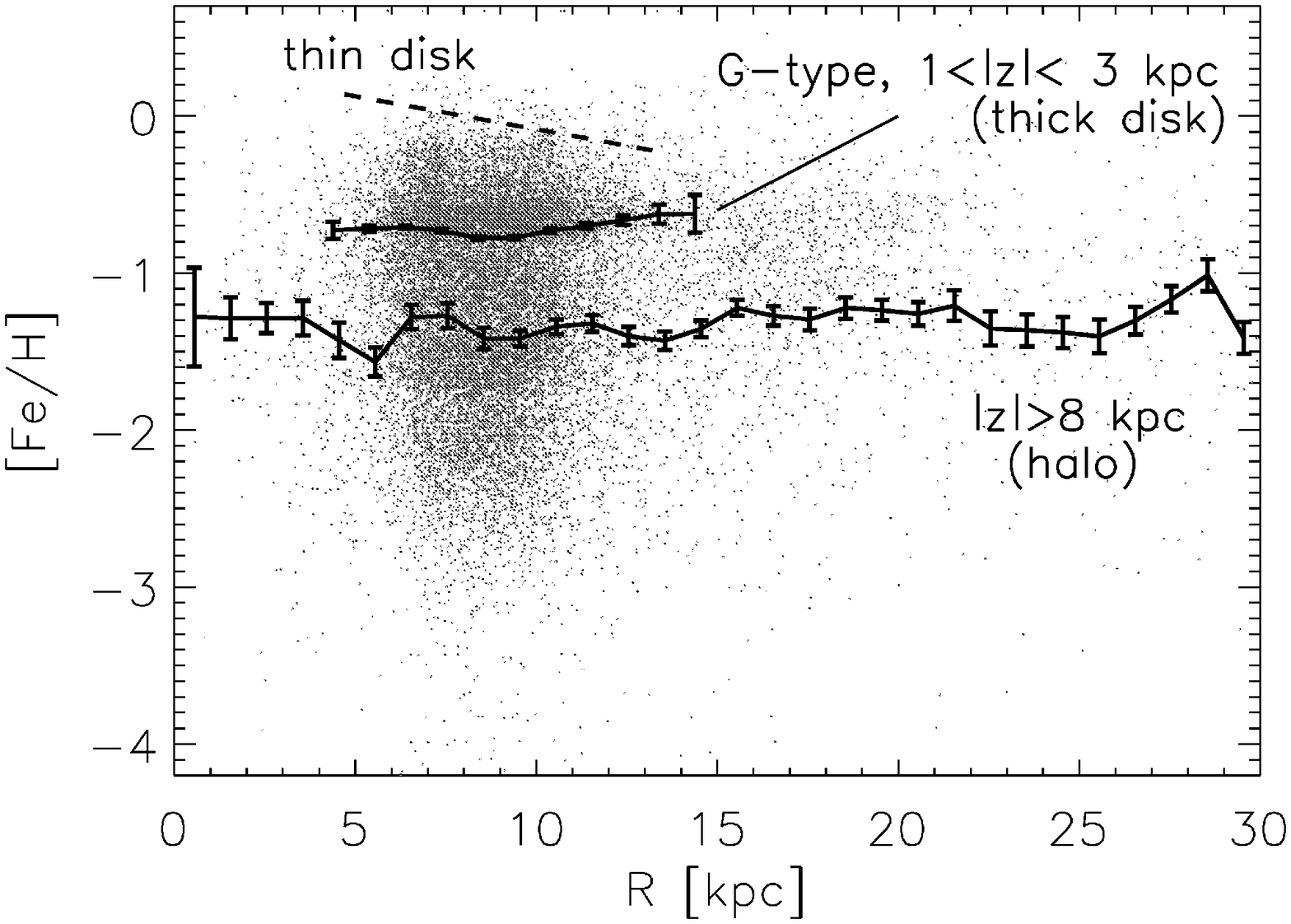} 
\caption{Metallicity distribution of the stars in the sample as a function of
galactocentric distance on the plane. 
All of the stars with $5000 <T_{\rm eff}< 7000$ K are
shown with gray dots. The solid lines correspond to the median values in 
1 kpc bins for stars more than 
8 kpc from the galactic plane (here tentatively 
identified as members of the halo), and to those with
 $5000 <T_{\rm eff}< 6000$ K located between 1 and 3 kpc from the plane
(a sample presumably dominated by thick-disk stars).
\label{rgal}}
\end{figure}

The frequency of spectroscopic binaries with periods shorter than 10 yr
among nearby (thin-disk) field stars is about 14 \% (Halbwachs et al. 2003).
Those with a mass ratio $q<0.5$ are most likely to be mistaken as single stars
in our analysis with no consequences, but those with $q > 0.8$, which
may represent as many as half of all binaries, have the potential to 
distort somewhat our results. Twin systems of F- or G-type stars 
will be confused by our procedures with single objects with lower
effective temperatures (by 100-200 K) and lower  
metallicities (by $\sim 1$ dex); 
albeit the H$\beta$ profile would be poorly matched.
Because our metallicity distribution for F-type stars is composed of 
similar numbers of stars at [Fe/H] $=-1.5$ and $=-0.7$, this bias is 
unlikely to be significant. The contamination could be larger for
G-type dwarfs, and deserves more detailed consideration in the future.

The thick-disk metallicity distribution seems to end abruptly at about solar
metallicity, but this result 
is affected by the bias against high-metallicity stars
imposed by the color selection of the SDSS spectrophotometric and reddening
standards. The contribution of thin-disk stars close to the plane 
to the solid line in
Fig. \ref{mdf}d, is negligible, as the shape of the distribution remains
essentially undisturbed when $|z|$ is restricted to the range $2 < |z| < 3$ kpc.
Again, changes are hardly noticeable if we extend the range to $0 < |z| < 3$
kpc. Most of the low-metallicity G-type stars near the plane exhibit halo
kinematics (see \S \ref{kin}). Close examination of the G-type dwarfs with
metallicities [Fe/H] $> -1.2$ within the range $1 < |z| < 3$ kpc reveals a mean
metallicity [Fe/H] $= -0.685 \pm 0.004$, a median value of [Fe/H] $ =-0.679$,
and a 1$\sigma$ dispersion of 0.238 dex. No vertical metallicity gradient is
apparent in this well-sampled region of the thick disk. If such a gradient is
present, it must be smaller than 0.03 dex kpc$^{-1}$.

The metallicity peak associated with F-type halo stars is slightly
shifted to higher metallicities at larger distances from the plane 
(See Fig. \ref{mdf}b). Because of the selection effects involved,  this
shift cannot be interpreted as a metallicity gradient in the halo. 
Thick-disk stars at $|z| < 3$ kpc might shift the peak of the
halo metallicity distribution to slightly higher values, and so 
potentially induce a metallicity shift in the opposite sense to
what is observed. On the other hand, constant magnitude limits might  
easily cause a metallicity bias in the sense that we observe. 
The absolute magnitude
of a mid-F dwarf is about $M_g= 4$ at solar metallicity, and 1 mag
fainter at [Fe/H]$=-1.7$. For the limits of the SDSS  
standards previously discussed, namely $15.5 < g < 18.5$, 
we are biased against high-metallicity F dwarfs at distances 
$d < 4$ kpc and against low-metallicity objects at $d > 5 $ kpc.

An enhancement is apparent in the metallicity distribution of distant F-type
stars at [Fe/H] $\sim -2.9$. No spatial or kinematic coherence is apparent for
these stars. Phenomenological models of galaxy formation in the framework of 
cold dark matter predict a similar feature in connection with the abrupt end of
the infall phase (Qian \& Wasserburg 2004). For halos originated from 2$\sigma$
density fluctuations the interpretation of the peak at this low metallicity
would be that infall cessation took place almost simultaneously with the start
of astration, about 0.5 Gyr after the Big Bang. 
Because of the uncertainties involved
due to small number statistics, more data are necessary to confirm the reality
of this feature in the metallicity distribution.

\begin{figure}[t!]
\includegraphics[angle=90,width=8cm]{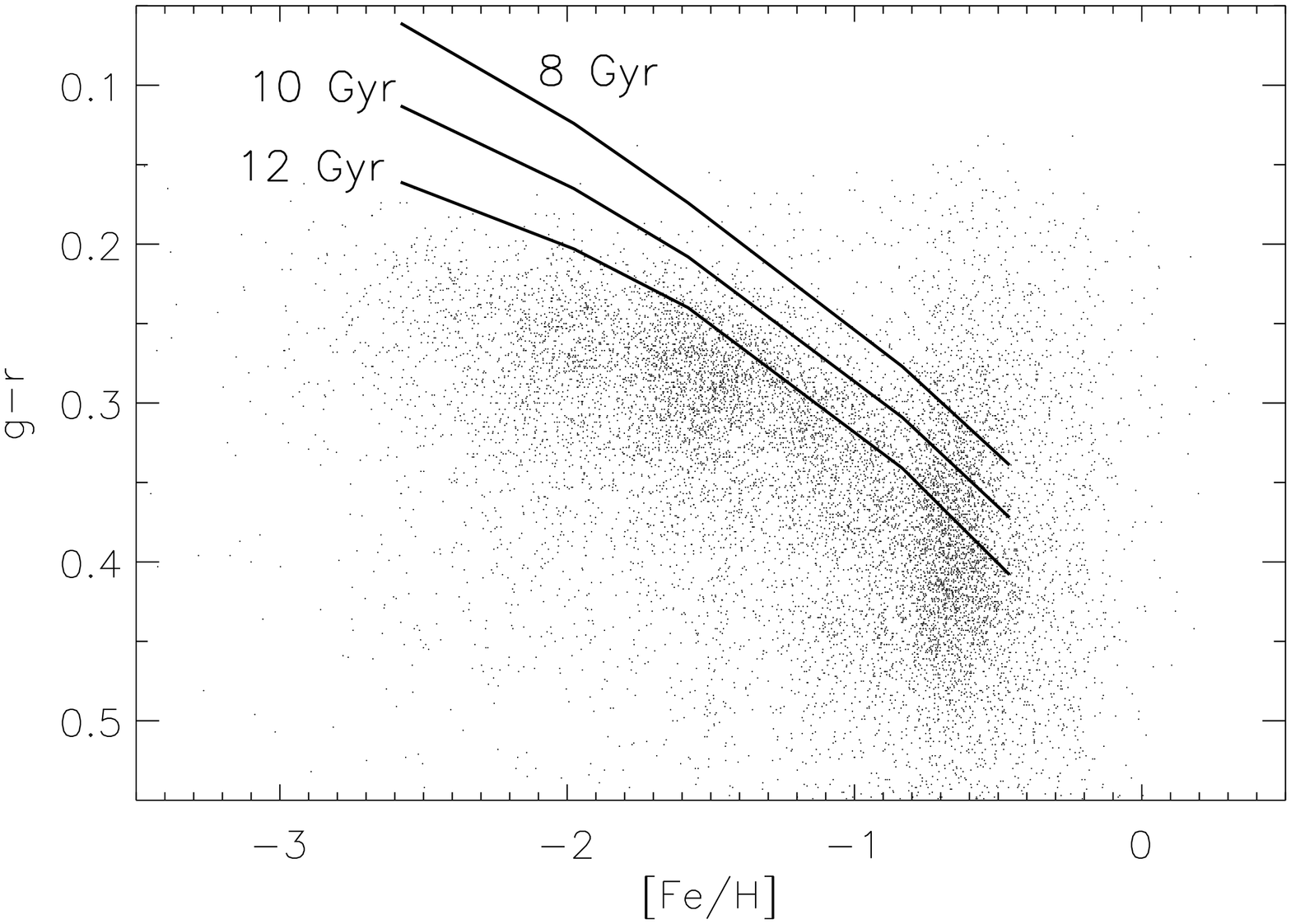} 
\caption{The color index $g-r$ (corrected for reddening) and metallicities
for F- and G-type dwarfs and subgiants ($\log g>4$) in the range $1<|z|<4$.
The solid lines trace the position of the main-sequence turnoff 
as a function of metallicity for ages 8, 10, and 12 Gyr.
\label{young}}
\end{figure}

Fig. \ref{rgal} shows the variations of stellar metallicities as a function
of the galactocentric distance projected onto the plane, $R$ (see Eq. \ref{rgaleq}). The
solid lines track the median metallicity in 1 kpc bins for stars at $|z|> 8$
kpc, and for those with $5000 < T_{\rm eff}<6000 $ K at $1< |z|< 3$. Halo stars
($|z|> 8$ kpc) exhibit a metallicity distribution that is essentially
independent of galactocentric distance. The thick-disk population, identified
with the G-type stars closer to the plane, can only be traced up to 
$\sim \pm 4$kpc from $R_{\odot}$. 
These stars also exhibit a flat or very small metallicity
variation, in contrast with the marked gradient found in the thin disk from open
clusters, Cepheids, H II regions, or massive stars (e.g., Twarog, Ashman, \&
Anthony-Twarog 1997; Andrievsky et al. 2004; V\'{\i}lchez \& Esteban 1996). The
dashed line in Fig. \ref{rgal} illustrates the average metallicity gradient in
thin-disk OB-type stars derived by Daflon \& Cunha (2004) over the range in
galactocentric distance spanned by their study. 

We note that the metallicity gradient of the thin disk may disappear 
beyond $R \sim 10-12$ kpc (Twarog et al. 1997; Yong et al. 2005), and 
in that regime the observed
abundance ratios show a more complicated behavior than what is observed in the
solar neighborhood (Yong et al. 2005). By comparing radial abundance gradients
measured from planetary nebulae, open clusters, cepheids, H II regions, and OB
stars and associations, Maciel, Lago \& Costa (2005) conclude that the thin-disk
gradient is now more flat by a factor of 2--3 compared with its value 8 Gyr ago
-- when star formation in the thick disk may have ended.

Our analysis suggests that in the temperature range $5000< T_{\rm eff}<7000$ K,
our sample contains over 150 stars with [Fe/H]$<-3$, and over 2000 stars with
[Fe/H]$<-2$. A few stars with [Fe/H] $< -4$ may be present in this sample, but
additional data are required to confirm their extreme metallicities. 
 
\subsection{Ages}

The changes between the metallicity distributions for $1<|z|< 3$ kpc 
in panels b
(F-type stars) and d (G-type stars) of Figure \ref{mdf} indicate that
in this range of $|z|$ there are both halo and thick-disk 
populations present among the F-type stars, but thick-disk members dominate the
sample of G-type stars. The SDSS targets metal-poor F-type stars 
for calibration purposes, but in
this range of $|z|, $ thick-disk stars should overwhelm the halo population, as it
is apparent among G-type stars. The presence of large numbers of halo F-type
dwarfs close to the disk is not the result of halo stars being younger than the
thick-disk population. It is instead a natural reflection of the fact that
metal-poor stars, despite living slightly shorter lives, have significantly
warmer surface temperatures. For any given age, the main-sequence turn-off 
is located at higher $T_{\rm eff}$s at [Fe/H]=$-1.4$ than at
[Fe/H]=$-0.7$. The bias against high-metallicity stars induced by the
color-based selection of the SDSS stellar targets is more severe at redder
colors (i.e., our G-types), and therefore must be a minor contributor. 
We can be more quantitative below, with the caveat that absolute ages 
are more difficult to estimate than relative numbers, and are subject 
to larger systematic errors.

Fig. \ref{young} shows the $(g-r)$\footnote{As always in this paper, we refer to
photometry corrected for reddening (see \S \ref{pre}).} index as a function of
the derived metallicity for the stars in the range $1<|z|<3$ kpc for which we
derived a high value of the gravity $\log g>4$. The imposed distance selection
automatically removes giants, and our gravity selection ensures such an
exclusion. Similar diagrams have been previously discussed by Gilmore, Wyse \&
Kuijken (1989). The turnoff stars define an upper envelope in this diagram,
which traces the oldest populations at any given metallicity. Using isochrones
from Girardi et al. (2004), we have determined the loci of the turnoff as a
function of metallicity for different ages, accounting for the enhanced
[$\alpha$/Fe] ratios using Eq. \ref{enhance}. We note that the predicted surface
temperatures for turnoff stars in the Yonsei-Yale isochrones (Yi, Kim \&
Demarque 2003) are very similar to those used here. The stars with [Fe/H]$<-1$
suggest that star formation in the halo ceased between 10 and 12 Gyr ago.

In general, very old ages, usually $>$ 10 Gyr, have been reported 
for thick-disk stars (see, e.g., Carney, Latham, \& Laird 1989; 
Quillen \& Garnett 2001; Reddy et al. 2003, 2005). Fig. \ref{young} does
not contradict those findings, but strongly supports that some members
of the thick-disk are at least as young as 8 Gyr.

\subsection{Kinematics}
\label{kin}

Additional insight on the nature of the 
stellar populations in the DR3 spectroscopic sample
may be obtained by exploiting the information provided by radial velocities. An
issue of particular interest is the rotation speed of the different galactic
components. Assuming any given star is simply moving in a circular orbit around
the Galaxy with a velocity V$_{\rm rot}$ in a cylindrical coordinate system
with its origin at
the center of the Galaxy $(R,\phi,z)$, we have 
{\bf V} = $-{\rm V}_{\rm rot}{\bf u_{\phi}}$, 
where ${\bf u_{\phi}}$ is the unit vector in the azimuthal
direction. Further assuming that the Sun is on a circular orbit 
in the galactic plane, with a velocity 
${\bf V_{\odot}} = -V_{\odot} {\bf u_{\phi}}$, 
V$_{\rm rot}$ can be inferred from the star's
position and the measured heliocentric radial velocity, $V_r$:

\begin{eqnarray}
\displaystyle {\rm V}_{\rm rot} & =  & 
\displaystyle \frac{V_{\odot} R \sin \phi - V_r d}
{R_{\odot} \sin \phi}  \nonumber \\
				&	       \nonumber \\
& = & \displaystyle  \frac{R}{R_{\odot}}  ~~ 
\left(V_{\odot} + \frac{V_r}{\cos b \sin l} \right).
\label{e1}
\end{eqnarray}

We avoid the approximation that the Sun is in a circular orbit by replacing, in
Eq. \ref{e1}, $V_{\odot}$ by the velocity of the local standard of rest (LSR;
adopted as 220 km s$^{-1}$, Kerr \& Lynden-Bell 1986), and $V_r$ by the radial
velocity with respect to the LSR
\begin{equation}
V'_r = V_r + U_{\odot} \cos b \cos l + V_{\odot} \cos b \sin l 
+ W_{\odot} \sin b,
\end{equation}
\noindent where the solar peculiar motion relative to the LSR is
$(U_{\odot},V_{\odot},W_{\odot}) = 
(10.1, 4.0 , 6.7)$ 
(Hogg et al. 2005).

\begin{figure}[t!]
\includegraphics[angle=0,width=8cm]{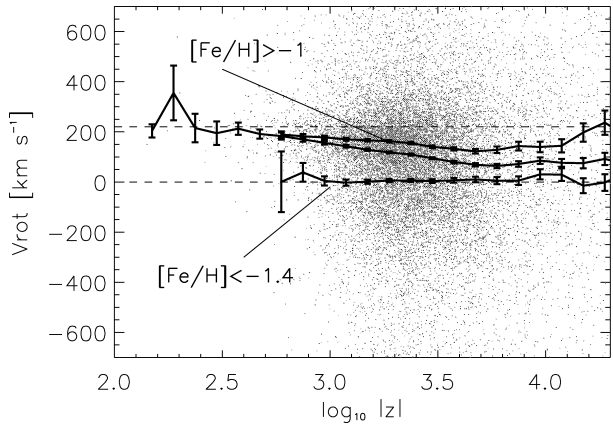}
\caption{Rotational velocity for individual stars derived assuming solid-rigid
rotation as a function of the stars' distance to the galactic plane.
The lines show the median value in non-overlapping 0.1 dex bins for all
stars, and two subgroups separated by metallicity. The
dashed lines mark the canonical values for the LSR (220 km s$^{-1}$) 
and no rotation.
 \label{vrot_z}}
\end{figure}

Fig. \ref{vrot_z} shows the derived V$_{\rm rot}$ as a function of the distance
to the plane for our sample. This transformation is appropriate only for stars
that have a purely azimuthal velocity. Therefore, it is a good approximation
only for thin-disk stars, as reflected by the small intrinsic scatter close to
the galactic plane. However, the median V$_{\rm rot}$ values for any given
population should be robust, and therefore statistically meaningful (for
reasonable values of $b$), as far as the Milky Way is approximately
axisymmetric. Fig. \ref{vrot_z} shows the median rotation velocity derived for
all stars, and two subgroups, in 0.1 dex bins. (Error bars are standard errors
for a mean value assuming a Normal distribution). This exercise reveals a smooth
variation of the median rotation velocities. The nearest stars approach the
thin-disk rotation, while, as the distance from the plane increases, the
inferred rotation velocity decreases up to a distance of 4 kpc. A vertical
asymmetry is clearly present in the region where most of the stars concentrate,
between 1 and 3 kpc (3 $< \log_{10} |z| \lesssim 3.5$). 
The lower solid line shows the
median values for stars with a metallicity [Fe/H]$< -1.4$, and confirms that the
galactic halo exhibits essentially no rotation. The upper line corresponds to
stars with [Fe/H]$>-1$; it should be dominated by the thick-disk population at
$1 < |z| < 3$ kpc, but is contaminated by halo stars at larger
distances from the plane. 

Based on the same model described in \S \ref{metal}, up to 18 \% of our
sample in the range $1<|z|< 3$ would be part of the stellar halo, while 
none of the stars would belong to the halo with a more restrictive window
between $1<|z|<2.4$ kpc. These figures are based on star counts alone. Selecting
only the stars with [Fe/H]$>-1$ should practically remove all halo stars 
in the range $1<|z|<3$ kpc. When this is done, a vertical gradient in the
derived median rotational velocity for thick-disk stars is apparent:
V$_{\rm rot} = 192 (\pm 8) - 0.016 (\pm 0.004)|z|$ km s$^{-1}$. 
This gradient was determined using all F- and G-type stars, but we obtain a   
result that is statistically indistinguishable 
when only G-type stars are used (ensuring a more complete removal
 of halo stars). Our analysis, thus, lends support
to previous measurements of a gradient by Majewski (1992), who derived 
$-21$ km s$^{-1}$  kpc $^{-1}$, and the reanalysis of his data by
Chen (1997; $-14 \pm 5$ km s$^{-1}$ kpc $^{-1}$),
although we derive a flatter slope that the determination by 
Chiba \& Beers (2000), who found $-30 \pm 3$ km s$^{-1}$ kpc $^{-1}$.

\begin{figure}
\includegraphics[angle=0,width=8cm]{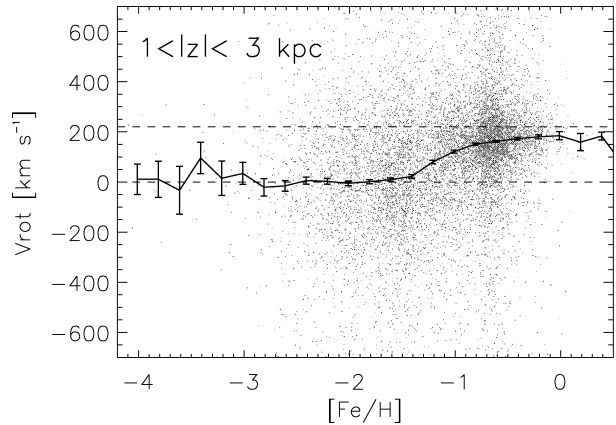}
\caption{Rotational velocity for individual stars at $|z|< 3$ kpc
derived assuming solid-body rotation.
The solid line shows the median value in non-overlapping 0.2 dex bins 
in metallicity. The
dashed lines mark the canonical values for the Sun (220 km s$^{-1}$) 
and no rotation.
 \label{vrot_feh}}
\end{figure}

A more detailed dissection of the dependence of the estimated galactic rotation
velocities on metallicity close to the plane ($1<|z|< 3$ kpc) is shown in Fig.
\ref{vrot_feh}. The solid line joins again the median values for the different
bins. This line remains essentially unchanged if the graph is restricted to
G-type stars. Thick-disk stars, here identified as moderately metal-poor
objects, lag the thin disk rotation by 0--100 km s$^{-1}$, with a median value
of $157 \pm 4$ km s$^{-1}$ at [Fe/H]=$-0.7$. 
In Fig. \ref{vrot_feh}, the
stars with $-3<$ [Fe/H] $<-1.4$ exhibit a flat rotation profile; the majority of
these objects are most likely halo members, with very limited contamination from
possible members of the metal-weak thick-disk population. The average value for
stars located at $|z|> 8$ kpc with metallicities [Fe/H]$<-1.4$ is 
V$_{\rm rot} = -25 \pm 15$ km s$^{-1}$,
a result that appears independent from the distance to the plane, as the median
values in Fig. \ref{vrot_z} suggest.

\section{Discussion and summary}
\label{sum}

We have analyzed a large number of stellar spectra included in the SDSS third
data release (DR3). By extensive comparison with modern spectroscopic libraries
and catalogs of atmospheric parameters, we have shown that our analysis
procedure is capable of deriving effective temperatures, surface gravities, 
and iron abundances with a $1\sigma$ accuracy of 3 \%, $0.3-0.4$ dex, 
and 0.2 dex,respectively, 
based on SDSS photometry and spectroscopy. We estimate a relative
precision of 20 \% in the inferred distances to dwarfs and subgiants, 
with systematic errors of similar size.

As a result of the selection of bright $0.2<(g-r)<0.4$ stars to calibrate SDSS
spectra, significant numbers of F-type subdwarfs are included in the 
DR3 spectroscopic sample.Many moderately metal-poor ($-1<$[Fe/H]$<-0.4$) 
late-F and early G-type dwarfs and subgiants are also present in the sample. 
We associate the observed distributions of distant metal-poor stars 
that show small galactic
rotation velocities with the halo population, and the closer V$_{\rm rot} \sim
160$ km s$^{-1}$ G-type stars of intermediate metallicity with the thick disk,
with the caveat that selection effects may make the observed distributions in
space and velocity significantly different from the true volume-limited
distributions.

The halo population shows a broad distribution of iron abundances 
which peaks at about [Fe/H] $ = -1.4$,
while the thick-disk stars exhibit a much narrower distribution 
with a sharp maximum near [Fe/H] $ = -0.7$. 
Halo stars follow a metallicity distribution that appears independent of
galactocentric distance on the plane; the same is true for the thick-disk
population between 5 and 14 kpc, in contrast with the marked gradient 
typically found in the thin disk. No vertical metallicity gradient is
discernible in the thick disk between $1 < |z| < 3$ kpc.
A clear vertical gradient is, however, detected in the thick-disk asymmetric 
drift. A linear least-squares fit to the median galactic rotational 
velocities of putative thick-disk dwarfs between 1 and 3 kpc from the 
plane indicates a slope of $-16$ km s$^{-1}$ kpc$^{-1}$ and,
if the linear trend is extrapolated, a  
maximum rotation speed of 192 km s$^{-1}$ 
(or a lag of 28  km s$^{-1}$ from the LSR) at $z=0$.

F-type stars at distances from the plane $|z|< 3$ kpc exhibit galactic rotation
velocities that are much lower than for thick-disk stars with higher
metallicities, and are consistent with the majority of them being drawn from
the halo population. 
The presence of these stars is naturally explained due to their lower
metallicity compared to the thick-disk population, which results in warmer
surface temperatures for a given mass. The halo population appears to be older
than about 11 Gyr, while there are thick-disk stars which are at least 2 Gyr
younger.

Models for the formation of the thick disk must satisfy several constraints now
apparent in the observations: (1) metal abundance ratios that are distinct from
those of thin-disk stars, (2) the lack of metallicity gradients both 
in the radial
and the vertical directions, (3) a vertical gradient in the rotation velocity,
and (4) the existence of thick-disk stars that are about 8 Gyr old, 
and thus overlap in age with the oldest stars in the thin disk. 
Hierarchical merging in the
context of cold dark matter has been recently shown to be compatible with the
enhanced $\alpha$/Fe ratios and the lack of metallicity gradients (Brook et al.
2004, 2005), but it remains to be seen whether it is consistent with the
observed vertical variation of rotational velocity. Gilmore, Wyse \& Norris
(2002) argue that this is possible, as the debris from the merging satellites
will end up in higher, more slowly-rotating orbits than stars newly formed due to the
heating of the thin disk, but detailed simulations are necessary. 

Efforts to acquire and study more complete and substantially less-biased samples
with the (SDSS) ARC 2.5M telescope and spectrographs are underway. 
The Sloan Extension for
Galactic Understanding and Exploration (SEGUE) aims at expanding SDSS imaging
coverage, in particular at lower galactic latitudes, over the next three years
of operation. A substantial fraction of the 250,000 stars selected for
medium-resolution spectroscopy as part of SEGUE will be a homogeneous sample of
G-type dwarfs; their space densities, abundances, and kinematics will set
substantially tighter constraints on models of 
the structure and evolution of the Milky Way.

\acknowledgments

CAP is thankful NASA's support (NAG5-13057, NAG5-13147). HJN acknowledges
funding from the National Science Foundation (AST 03-07571). TCB and YSL
acknowledge partial support from grants AST 00-98508, AST 00-98549, AST
04-06784, and PHY 02-16783, Physics Frontier Centers/JINA: Joint Institute for
Nuclear Astrophysics, awarded by the US National Science Foundation. This
paper has greatly benefited from discussions with, and comments from, Johan
Holmberg, David Lambert, Iv\'an Ram\'{\i}rez, and the referee, Bruce Carney. 
We are indebted to Ivan Hubeny and Thierry Lanz for making their codes 
publicly available.
Funding for the
creation and distribution of the SDSS Archive has been provided by the Alfred P.
Sloan Foundation, the Participating Institutions, the National Aeronautics and
Space Administration, the National Science Foundation, the U.S. Department of
Energy, the Japanese Monbukagakusho, and the Max Planck Society. The SDSS Web
site is http://www.sdss.org/.

The SDSS is managed by the Astrophysical Research Consortium (ARC) 
for the Participating Institutions: 
The University of Chicago, Fermilab, the Institute for Advanced Study, 
the Japan Participation Group, The Johns Hopkins University, 
the Korean Scientist Group, Los Alamos National Laboratory, 
the Max-Planck-Institute for Astronomy (MPIA), 
the Max-Planck-Institute for Astrophysics (MPA), 
New Mexico State University, University of Pittsburgh, 
Princeton University, the United States Naval Observatory, 
and the University of Washington.

\end{document}